\documentclass[a4paper,12pt]{article}

\usepackage[a4paper,top=2cm,bottom=2cm,left=3cm,right=2cm]{geometry}
\usepackage{amssymb,graphicx}
\usepackage{epstopdf}
\usepackage{amsmath,amsfonts}
\usepackage{mathtools,color}
\usepackage{hyperref}
\usepackage{caption}
\usepackage{subcaption}
\RequirePackage[numbers,sort&compress]{natbib}

\def\d{\partial}

\newcommand{\be}{\begin{equation}}
\newcommand{\ee}{\end{equation}}
\newcommand{\bea}{\begin{eqnarray}}
\newcommand{\eea}{\end{eqnarray}}
\newcommand{\bg}{\begin{gather}}
\newcommand{\eg}{\end{gather}}
\newcommand{\bseq}{\begin{subequations}}
\newcommand{\eseq}{\end{subequations}}

\def\half{\frac{1}{2}}

\newcommand{\pd}[2]{\ensuremath{\frac{\d #1}{\d #2}} }
\newcommand{\vev}[1]{\left\langle #1\right\rangle}

\DeclareMathOperator{\tr}{tr}

\newcommand{\pt}{\partial}

\newcommand{\I}{\mathcal I}

\renewcommand{\L}{\mathcal L}

\numberwithin{equation}{section}

\begin{document}

\vspace{.5cm}

\begin{center}
{\Large\sc Solid Holography and Massive Gravity}

\vspace{1.cm}

\textbf{Lasma Alberte$^a$, Matteo Baggioli$^{b,c}$, Andrei Khmelnitsky$^a$, Oriol Pujol{\`a}s$^b$
}\\

\vspace{1.cm}
${}^a\!\!$ {\em {Abdus Salam International Centre for Theoretical Physics\\Strada Costiera 11, 34151, Trieste, Italy}}

\vspace{.2cm}
${}^b\!\!$ {\em {Institut de F\'isica d'Altes Energies (IFAE)\\ 
The Barcelona Institute of Science and Technology (BIST)\\
Campus UAB, 08193 Bellaterra (Barcelona) Spain
}}

\vspace{.2cm}
${}^c\!\!$ {\em {
Department of Physics, Institute for Condensed Matter Theory\\
University of Illinois, 1110 W. Green Street, Urbana, IL 61801, USA
}}

\end{center}

\vspace{0.8cm}

\centerline{\bf Abstract}
\vspace{2 mm}
\begin{quote}\small

Momentum dissipation is an important ingredient in condensed matter physics that requires a translation breaking sector. In the bottom-up gauge/gravity duality, this implies that the gravity dual is massive. 
We start here a systematic analysis of holographic massive gravity (HMG) theories, which admit field theory dual interpretations and which, therefore, might store interesting condensed matter applications.
We show that there are many phases of HMG that are fully consistent effective field theories and which have been left overlooked in the literature. 
The most important distinction between the different HMG phases is that they can be clearly separated into {\em solids} and {\em fluids}. This can be done both at the level of the unbroken spacetime symmetries as well as concerning the {\em elastic} properties of the dual materials. We extract the {\em modulus of rigidity} of the solid HMG black brane solutions and show how it relates to the graviton mass term. We also consider the implications of the different HMGs on the electric response. We show that the types of response that can be consistently described within this framework is much wider than what is captured by the narrow class of models mostly considered so far.

 \end{quote}

\vfill
 
\newpage

\tableofcontents

\newpage

\section{Introduction and Motivation}

Recently, a somewhat surprising connection between Condensed Matter (CM) and Massive Gravity (MG) has been unveiled: MG theories are {\em especially} relevant for CM applications by way of the AdS/CFT correspondence~\cite{Vegh:2013sk,Blake:2013bqa,Blake:2013owa,Davison:2013jba}. This 
is quite appealing 
because CM contains a large variety
of poorly understood systems, 
whereas both MG and AdS/CFT have enjoyed a vast development in the last decade.

The  fact that  CM and MG must be connected somehow is actually very natural: any distribution of matter gives rise to a plasma mass for the graviton, much in the same way the photon gets a mass in a medium with a finite density of freely moving charges. Since a plasma mass is linked to the presence of a material, it is inevitably non-Lorentz invariant and so it comes as no surprise that it can be connected with Lorentz Violating Massive Gravity (LVMG). 
It is well known now that there are many types or phases of LVMG  that are free from the consistency problems which plague the Lorentz invariant versions of massive gravity~\cite{Rubakov:2004eb,Dubovsky:2004sg}. Moreover, the LVMGs are believed to be better behaved also in the UV region in comparison to the LI massive gravity theories.\footnote{Recently it has been shown that some forms of  LVMG can even be UV-completed  to almost Planckian energy scales~\cite{Blas:2014ira}.}
Thus, it seems to be possible to match the great diversity of types of condensed matter materials with the great diversity of consistent phases of LVMG.

This connection, surely quite obvious to most MG experts, has always been clear but seemed of little consequence  
because the graviton plasma mass inside a material with an energy density $\rho$ 
can be estimated as $m_g^2 \sim \rho\; M_{\text{Pl}}^{-2}$, which is much smaller than the inverse size of  human-scale materials. 
Cosmology is the area where one can have, in principle, more significant effects, even if to date it is unclear whether any form of MG may help with the open problems in cosmology (for a recent review see, e.g.~\cite{deRham:2014zqa}).

In AdS/CFT correspondence~\cite{Maldacena:1997re}, though, this connection opens an entirely new dimension. MGs in AdS-like spaces are dual to strongly coupled materials that incorporate a crucial aspect of CM: a sector that breaks spatial translation invariance
. From the CM perspective, including this sector in the low-energy description is of central importance to capture {\em e.g.} the effects from phonons and from disorder (see below). 

In our view, however, the way how MG plays this central role
has not been clearly stated yet in the AdS/CMT literature. To make contact with the AdS/CMT applications, we shall stick to the common terminology that refers to the MG theories in AdS (or more general spaces that allow a holographic interpretation) as {\em Holographic Massive Gravity (HMG)}. 
The main aim of this paper is to clarify the role, consistency, and classification of HMG theories. The key messages will be:
\begin{enumerate}
\item HMGs can be broadly and clearly separated into solids and fluids\footnote{A more thorough classification of the possible phases of HMG including  superfluid relatives as well as other distinctions along the holographic direction is deferred for future work
.}. This follows both from the analysis of the residual diffeomorphisms that are preserved as well as from the presence/absence of an {\em elastic shear response}.

\item The large family of HMGs considered below are very healthy: they are fully consistent as Effective Field Theories (EFTs) in that they are free of ghosts and other pathologies at linear and nonlinear level, even if they are not necessarily of dRGT type~\cite{deRham:2010kj}. 
In addition, and in contrast to the application to cosmology, in holography the graviton mass does not need to be so much smaller than the Planck mass: it suffices that it is around the AdS curvature scale $1/L$ wich can be taken to be, say, one to two orders of magnitude below $M_{\text{Pl}}$. In this case, the strong coupling scale (which is usually a geometric mean of the form $\Lambda_n\sim(m_g^{n-1} M_{\text{Pl}})^{1/n}$ for $n=2,3,5$) associated with the graviton mass sector does not differ significantly from the Planck scale, and in any case is not below $1/L$. This is quite important for the holographic application because it allows the  gravitational description to continue to be weakly coupled (at least at low frequencies and long wavelengths) even in the presence of a fully nonlinear mass sector.

\item The class of HMGs relevant for CM is much more general than the dRGT models~\cite{deRham:2010kj} that have been considered mostly so far (see {\em e.g.}~\cite{Vegh:2013sk,Blake:2013bqa,Davison:2013jba}), as was already argued in ~\cite{Baggioli:2014roa}. The family of theories that are free from various pathologies (as discussed {\em e.g.} in~\cite{Baggioli:2014roa,Alberte:2014bua}) and have the same symmetries is parameterized by free functions that are related to the physical properties of the materials. 
This has some practical implications because there are many properties of the dRGT-like massive gravities that are not  generic in the AdS/CMT context, and therefore can lead to misguided implications for CM. For instance, the full set of HMG theories that can be applied to modelling the low energy limit of CM systems is not only parameterized by two real numbers.  
\end{enumerate}
All in all, this suggests that holographic massive gravities might be very relevant for condensed matter applications.

In this paper we shall  use both the presentation of massive gravity in terms of broken diffeomorphisms and in its  covariantized form relying on the St\"uckelberg fields $\phi^A$. In the latter language MG can be seen as a theory of general relativity coupled to a number of scalar fields. The application of the scalar fields formalism to holographic massive gravity was initiated in~\cite{Andrade:2013gsa,Taylor:2014tka} and was instrumental in identifying the additional degrees of freedom as phonons~\cite{Baggioli:2014roa}.

 The most meaningful way to characterize the various different phases of massive gravity is phrased in terms of the symmetries that are left unbroken in each phase. The situation is very similar to the 
organization of the 
EFTs for various materials according to the spontaneous symmetry breaking pattern of the time and/or space translations,~\cite{Leutwyler:1993gf,Leutwyler:1996er,Nicolis:2015sra,Nicolis:2013lma}. 
The classification of the phases of MG proceeds in very similar terms because on physical grounds the LV mass terms are the possible forms of the plasma-mass that is generated in different types of materials.  In other words, the EFTs of solids/fluids/etc. are related to the phases of LVMG by gauging the spacetime symmetries, that is, by introducing the coupling to the dynamical metric by the usual covariantization prescription. Thus it is not at all surprising that one can speak of, {\em e.g.}, solid and fluid phases of MG. For the sake of simplicity, we shall restrict ourselves to these two cases and defer for future work a more thorough analysis of other phases of MG.

As is nowadays relatively well-understood, the EFT for  fluids and solids  in flat space involves a set of phonon scalars $\phi^I$ (in 2+1 dimensions,  $I=1,2$) that enjoy internal shift and rotation symmetries for homogeneous and isotropic materials~\cite{Leutwyler:1993gf,Leutwyler:1996er,Nicolis:2015sra,Nicolis:2013lma}. The internal symmetry group for solids is the two-dimensional Euclidean group of translations and rotations. For fluids, the internal group is much bigger and includes also volume preserving  diffeomorphisms (VPDiffs).
The scalars acquire an expectation value $\vev{\phi^I}=\delta^I_i x^i$ and break the product of the (space  transformations) $\otimes$ (internal transformations) to the diagonal subgroup. For fluids, the preserved symmetry includes a volume preserving diagonal subgroup. 

The effective Lagrangian at the lowest order in derivatives in the two cases can be written as~\cite{Dubovsky:2011sj,Son:2005ak,Endlich:2012pz,Nicolis:2013lma}
\be\label{Lsf}
L^{(solids)}=V_s(X,Z) \qquad {\rm and} \qquad L^{(fluids)}=V_f(Z) \;,
\ee
where $X=\tr \I^{IJ}$ and $Z=\det \I^{IJ}$ with $\I^{IJ}\equiv \eta^{\mu\nu}\pt_\mu\phi^I\pt_\nu\phi^J
$.\footnote{In higher dimensions there are more invariants. For instance, in 3+1 dimensions $\tr \I^{I}_{J}\I_{K}^{J}$ gives an independent invariant.} 
The functions $V_{s,f}$ encode the linear and nonlinear properties of the solid and fluid, and they are free functions subject to mild consistency constraints. 
It is easy to realize that gauging these theories leads to graviton mass terms  around the solution with 
$\vev{\phi^I}=\delta^I_i x^i$. The simplest way to see this is to replace $\eta_{\mu\nu}$ with $g_{\mu\nu}$ and to go to the unitary gauge where the scalar fields are fixed to be equal to their background configuration. The above solid/fluid Lagrangians then become nonlinear potential terms for the metric
\be\label{Vsf}
V_s\big(\,\tr g^{ij},\, \det g^{ij}\,\big)  \qquad {\rm and} \qquad V_f\big(\, \det g^{ij}\,\big)~,
\ee
where $g^{ij}$ denotes the spatial part of the inverse metric. We note that exactly the same procedure  was followed in the so-called `solid inflation', ref.~\cite{Endlich:2012pz}.

At this point, it is quite clear that there must exist a fully equivalent nonlinear formulation of solid/fluid MGs that is phrased entirely in terms of a unitary-gauge metric variable. The form of this action is dictated by requiring it to be invariant under certain subset of the diffeomorphisms, that do not include the spatial diffeomorphisms $x^i \mapsto \tilde x^i(t,x^j)$. The preserved diffeomorphisms are the ones enjoyed by the potential terms in \eqref{Vsf}. Both for solid and fluid MGs, these include the time-reparametrizations $t \mapsto f(t)$ 
  plus global translations and rotations  that force the potential not to depend explicitly on $x^i$ and to contract the spatial indices with Kronecker delta $\delta_{ij}$.\footnote{For simplicity, we shall assume  that he kinetic part of the action is given by the standard Einstein-Hilbert term, which is not the most general one compatible with these symmetries.}
For fluid MG the potential is also invariant under the spatial VPDiffs, that forces it to be a function of $\det{ g^{ij}}$ only. Importantly, as we shall see below, the VPDiff symmetry protects the vanishing of the physical mass parameter of the metric tensor modes.

In order to link the defining symmetries of solid/fluid MGs to the structure of mass parameters, it is important to identify the relevant notion of the mass terms, which is not entirely obvious since we need to work in curved backgrounds. The most practical definition is to expand  the action around the solutions of interest and look at the non-derivative quadratic terms once the kinetic terms have a canonical form. 
Limiting the discussion here to homogeneous and isotropic backgrounds, one can follow~\cite{Rubakov:2004eb,Dubovsky:2004sg} and parameterize the possible mass terms by five constant mass parameters,\footnote{The precise link between these mass terms and the nonlinear Lagrangian \eqref{Vsf} will be done in full detail in sections~\ref{sec:phases} and~\ref{sec:two_fields} in a class of models that covers fluids and solids.}
\be\label{masses}
m_0^2 h_{00}^2+2m_1^2 h_{0i}^2 - m_2^2 h_{ij}^2+ m_3^2 h_{ii}^2 -2 m_4^2 h_{00} h_{ii} \; .
\ee

For fluids/solids, the preserved symmetries at the level of the Lagrangian in the unitary gauge formulation, and the allowed mass terms are as follows:
\begin{enumerate}

\item[{\bf Solids}] Preserved symmetries: time reparametrizations and the diagonal subgroup of space translations and rotations. Expressed as infinitesimal diffeomorphisms ($x^\mu\to x^\mu+\xi^\mu(x^\nu)$), these are: $\big\{\, \xi^t(t)$,  $\xi_\oplus^i=c^i+\epsilon^i_jx^i \big\}$ where $c^i=const,\,\epsilon^T=-\epsilon$. The $\oplus$ label denotes that it is a combination of a spatial translation/rotation and a corresponding internal transformation that leaves the background configuration invariant.
The allowed mass terms are $m_{1,2,3}\neq0$

\item[{\bf Fluids}] Preserved symmetries: time reparametrizations and the diagonal subgroup of translations, rotations, and volume-preserving diffeomorphisms. In infinitesimal form, these look like $\big\{\,\xi^t(t)$,  $\xi^i_{VP\oplus}(x^j)\big\}$ (VP = volume preserving, $\partial_i\xi_i=0$).  The allowed mass terms are $m_{1,3}\neq0$. Importantly, for fluids $m_2=0$.

\end{enumerate}

Let us insist that in both cases the spatial translations are broken (or non-linearly realized), which we emphasize with the $\oplus$ label that should remind that this is not a standard translation. In the realization of these phases as HMGs presented in the main body of the paper, both cases lead to a finite DC conductivity in the electric response. 
Also, both the fluid/solid EFTs and the fluid/solid MGs are {\em very} healthy in the sense that the Lagrangian can be always chosen in order to avoid ghosts or other pathologies at linear and nonlinear levels. We show that explicitly in Section 5. 

The main differences between the two types of theories are i) that the solid phases exhibit propagating transverse phonons --- the Goldstone modes of the broken space translations, inhomogenenous in spatial coordinates --- whereas the fluids do not; and ii) that the tensor modes are massive/massless for solid/fluid  phases respectively. 
In the HMG constructions below, both statements will hold at the level of local propagating fields in AdS. However, since local degrees of freedom in the dual field theory (such as the transverse phonons) can be over-damped in black brane backgrounds, the difference in this respect between the solid/fluid HMGs becomes slightly blurred. Instead, a clear distinction survives at the level of the tensor modes being massive/massless. In turn, as we shall show below, this allows to 
characterize the solid types of HMGs as the ones with  \emph{rigidity}, a nonzero static elastic shear response, encoded in the non-vanishing $m_2$.\smallskip

After this quite generic review of the physical distinctions between MG phases, let us introduce a bit more the specifics of the holographic application  (for some recent developments concerning the importance of the translation breaking sectors see~\cite{Hartnoll:2009sz,Hartnoll:2008hs,Iqbal:2008by,Vegh:2013sk,Blake:2013bqa,Blake:2013owa,Davison:2013jba,Hartnoll:2009ns,Davison:2013txa,Hartnoll:2014lpa} and references therein). As already mentioned, we will focus on asymptotically $AdS_4$ backgrounds, since they admit simple and well-understood 2+1 CFT interpretations. We shall denote by $t$ and $x^i$ the coordinates along the boundary directions and $r$ the holographic coordinate dual to the renormalization group scale. Since the holographic map already gives $r$ a special role, we consider gravity theories  with an anisotropic mass term. The crucial ingredient to make contact with realistic CM is to break translations in the $x^i$ directions. Thus, in the non-unitary gauge description we are going to need at least two scalars $\phi^I$ for $I=1,2$ with vacuum expectation values $\vev{\phi^I}=\delta^I_i x^i$.\footnote{There are other ways to accomplish the breaking of $x^i$-translations like the so-called holographic lattices, {\em i.e.}, explicitly $x^i$-dependent source terms. For this purpose, these constructions are equivalent to massive gravity \cite{Blake:2013owa} and considerably less convenient to work with.} One can also consider additional radial and temporal fields $\phi^r$ and $\phi^t$. The role of a  $\phi^t\propto t$ scalar is to introduce {\em super}-fluid/solid versions of the dual material, and the role of $\phi^r$ is basically the same as of the commonly-used dilaton field. This identification especially holds when $\phi^r$ does not enjoy an internal translational symmetry, and thus provides a rich variety of radial dependence in black brane as well as in vacuum solutions (zero temperature and density). For the sake of simplicity, we shall restrict here mostly to the simplest case that provides momentum dissipation in the $x^i$ directions --- a two-field model with the $\phi^I$ scalars only.

These two-field HMG models are already quite rich (as they include solid and fluid behaviours), simple and interesting. Because of their dual field theory interpretation, it seems appropriate to call them {\em solid/fluid CFTs}. 
Indeed, the asymptotically AdS black brane solutions in the solid/fluid MG admit an interpretation as CFTs with solid/fluid behaviour that are able to dissipate momentum, and which are deformed only by a finite temperature and finite density. The fact that momentum is not conserved can be seen as a coupling of a standard momentum-conserving CFT to a translation-breaking sector.
The latter is realized in the black brane solutions via the $\phi^I$-hair. 
The magnitude of the translation-breaking is proportional to the local energy density in the $\phi^I$ sector and is position- ({\em i.e.}, scale-) dependent. Importantly, depending on the form of the potential $V(X,Z)$ this energy density can be located closer or farther from the black brane horizon. In the CFT, this characterizes the typical energy scale of the translation breaking. 
At least for the solids it seems suggestive and consistent to identify the {\em explicit} breaking (that is localized near the AdS boundary) as some type of \emph{lattice disorder} in the sense that it mimics a random distribution of lattice defects in an otherwise perfect crystal. 
In any case, the presence of defects seems unrelated to the fact that momentum dissipates, as is clear a posteriori since the DC conductivities involve horizon data only.
We shall not deepen much more into the nature of this disorder in this work  (for the role that holographic version of disorder can play in various contexts, see~\cite{Donos:2014uba,Donos:2014cya,Amoretti:2014ola,Grozdanov:2015qia,Davison:2014lua,Hartnoll:2014cua,Gouteraux:2014hca,Hartnoll:2015rza,Lucas:2015lna,Lucas:2014sba,Lucas:2014zea,Arean:2015sqa,Arean:2013mta,Amoretti:2015gna,Baggioli:2015dwa,Baggioli:2015zoa,Baggioli:2015gsa} and references therein). Still, having identified 
a clear distinction between solid and fluid materials might help identifying the nature of disorder, simply because the notion of disorder itself seems much better defined for solids than fluids in the first place.  

The focus of our work is to analyze the full set of mass terms that are available in HMG and to clarify how the various mass terms control different transport properties of the dual materials. Obviously, there are many more mass terms if one restricts to homogeneity and isotropy of the CFT in the spatial directions, $x^i$, but not in the holographic direction. Hence, the full classification of the phases of HMG is rather complicated. We shall therefore initiate the analysis of the effects from the full range of possible mass terms, but will also restrict to the cases that are simpler to interpret. Along the way, we will perform various checks such as that the DC conductivity is indeed finite in both solids and fluids, and that the solids, in addition, enjoy a nontrivial elastic response. 

The rest of this paper is organized as follows. In section~\ref{sec:setup} we set up our notations and introduce a classification of the massive gravity theories according to the preserved subgroup of the diffeomorphism symmetry. We describe two particular cases, on which we focus in the present paper: the general theory with negligible background stress energy tensor and the theory written in terms of two St\"uckelberg fields --- the simplest theory describing the solid phase of HMG. Section~\ref{sec:response} is devoted to the electric response in the dual field theory. We derive the expression for the DC conductivity in terms of the mass parameters and discuss the phenomenology of the representative models. We proceed in section~\ref{sec:elastic} with discussing the elastic properties of the duals to the solid and fluid phases of HMG. Section~\ref{sec:two_fields} contains the analysis of the massive gravity with two scalar fields. Section~\ref{sec:conclusions} is devoted to the conclusions and outlook. A detailed analysis of the HMGs with four scalar fields and zero stress-energy tensor is presented in appendices~\ref{sec:linear} and~\ref{sec:stab}.

%
%
%
%
%
%
%
%
%
%

\section{Phases of Holographic Massive Gravity}\label{sec:setup}
\label{sec:phases}

For holographic applications in condensed matter theory we are interested in massive gravity theories that allow for asymptotically AdS charged black brane solutions.  The action that will be considered in this paper is the Einstein-Maxwell action with a negative cosmological constant and a graviton mass term:\footnote{We work in the units where $M_{\text{Pl}}^2=(8\pi G)^{-1}\equiv 1$.} 
\be\label{action}
S=\int d^4x\sqrt{-g}\,\left[\frac{1}{2}\left(R+\frac{6}{L^2}\right)-\frac{L^2}{4}F_{\mu\nu}F^{\mu\nu}+\mathcal L_\phi\right] \; .
\ee
The graviton mass term $\mathcal L_\phi$ can be written as a Lagrangian for the St\"uckelberg scalar fields $\phi^A$ and will be specified in the following sections. We shall concentrate on the probe limit and neglect the backreaction from the mass term on the background metric everywhere apart from section~\ref{sec:two_fields}. In this limit the action admits an asymptotically AdS Reissner--Nordstr\"om black brane solution:
\be
\label{solmetric}
ds^2 = \hat g_{\mu\nu} dx^\mu dx^\nu =  L^2\left(\frac{dr^2}{f(r)r^2}+\frac{-f(r)dt^2+dx^2+dy^2}{r^2}\right) \;,
\ee
with the emblackening factor $f(r)$ given by
\be\label{fr}
f(r)=1-Mr^3+\frac{\mu^2}{2r_h^2}r^4 \; .
\ee
Here the mass $M$ is a constant of integration that can be determined by demanding that $f(r)$ vanishes on the horizon $r=r_h$ to be
\be
M=\frac{1}{r_h^3}+\frac{\mu^2}{2r_h} \; .
\ee
The solution for the St\"uckelberg scalars and the Maxwell field takes the form
\be\label{sol}
\hat \phi^A=x^\mu\delta^A_\mu \; ,	\qquad 	\hat A_t = \mu\left(1-\frac{r}{r_h}\right) \;.
\ee

In the following sections we shall discuss the different phenomenological consequences in the dual theory induced by the presence of the graviton mass in the bulk and investigate the bulk stability of the perturbations around the black brane background. To do so we consider the perturbations of the metric and Maxwell fields defined as
    \be
  g_{\mu\nu}=\hat g_{\mu\nu}+h_{\mu\nu} \; ,\qquad A_\mu=\hat A_\mu+a_\mu \; ,
  \ee
and the perturbations of the St\"uckelberg scalar fields $\pi^A$ defined as
\be\label{pi}
\phi^A=\hat\phi^A + \pi^A \; .
\ee

Since we are interested in homogenous and isotropic condensed matter systems, we shall limit ourselves to the mass terms that preserve the constant shifts and rotations of the transverse coordinates $x^i = \{x,y\}$. It allows us to classify the perturbations according to the scalar, vector, and tensor representations of the transverse $O(2)$ rotation group. The split depends on whether we consider homogeneous modes, i.e. independent on the transverse coordinates $x^i$, or inhomogeneous modes. For simplicity we shall consider only the homogeneous case when the perturbations can be classified as
  \begin{align}
  &\textrm{scalar:}\qquad h_{tt},\,h_{tr},\,h_{rr},\,h\equiv \frac{h_{ii}}{2},\,a_t,\,a_r,\,\pi^t,\,\pi^r \; ;\label{scalar}\\
  &\textrm{vector:}\qquad h_{ti},\,h_{ri},\,a_i,\,\pi^i \; ;\\ \label{tensor}
  &\textrm{tensor:}\qquad \bar h_{ij}\equiv h_{ij}-\frac{h_{kk}}{2}\delta_{ij}, \,\bar h_{ii}=0 \; .
  \end{align}

\subsection{General mass terms}\label{sec:masses}
A straightforward way of analyzing the stability and phenomenology of a general theory of massive gravity is to start with the non-covariant form of the massive gravity action with the mass term written it terms of the metric perturbation $h^{\mu\nu}$ and then restore the diffeomorphism invariance by the St\"uckelberg trick. A similar analysis of the stability of the Lorentz violating massive gravity around Minkowski background has been performed previously in refs.~\cite{Rubakov:2004eb,Dubovsky:2004sg}. 

We define the inverse metric perturbations\footnote{We choose to define the mass term in terms of the inverse metric perturbations due to the usual Lorentz invariant convention to write the graviton mass in terms of the matrix $g^{\mu\lambda}f_{\lambda\nu}$ where $f_{\mu\nu}$ is an auxiliary reference metric.} as
\be\label{mpert}
g^{\mu\nu}=\hat g^{\mu\nu}+ h^{\mu\nu}\equiv g^{\mu\nu}-\hat g^{\alpha\mu}\hat g^{\beta\nu}h_{\alpha\beta}+\mathcal O\left(h_{\mu\nu}^2\right) \; ,
\ee
and write the most general quadratic mass term that preserves the rotations of the transverse coordinates in the following form:
\begin{align}\label{acth}
\mathcal L_\phi (h^{\mu\nu},r)&= \frac{1}{2}\left ( m_{0}^2(r) (h^{t t})^2+ 2m_{1}^2(r) h^{t i} h^{t i } - m_2^2(r) h^{i j} h^{i j}  \right. \notag \\
&\quad  + m_{3}^2(r) h^{i i} h^{j j} -2m_{4}^2(r) h^{t t} h^{i i} \notag\\
&\quad  + m_{5}^2(r) (h^{r r})^2 + m_{6}^2(r) h^{t t} h^{r r }   \notag \\
&\quad   +m_{7}^2(r) h^{r i} h^{r i } + m_{8}^2(r) h^{t i} h^{r i } +m_{9}^2(r) h^{r r} h^{i i} \notag\\
&\quad + m_{10}^2(r)h^{rt}h^{rt} +m_{11}^2(r)h^{rt}h^{ii}+m_{12}^2(r)h^{tt}h^{rt}+m_{13}^2(r)h^{rr}h^{rt}\big ) \;,
\end{align}
where all the masses $m_i^2$ are functions of the radial coordinate. The repeated transverse coordinate indices are contracted with $\delta_{ij}$. The numbering of the mass parameters is chosen to be consistent with the notations of refs.~\cite{Rubakov:2004eb,Dubovsky:2004sg}. The novelty introduced by the planar AdS black brane background is the special role of the holographic coordinate $r$. For future reference we note that the mass parameters defined in \eqref{acth} can be classified with respect to the perturbations that they affect as:
    \begin{align}
  &\textrm{scalar:}\qquad  m_0,m_{2-6},m_{9-13}\;;\\
  &\textrm{vector:}\qquad  m_1,m_7,m_8\; ;\\ 
  &\textrm{tensor:}\qquad  m_2\; .
  \end{align}
We note that this classification is valid only for the homogeneous modes. 

\subsection{Symmetries}\label{sec:symmetries}
The mass term \eqref{acth} explicitly breaks the spacetime diffeomorphisms since the metric perturbation $h^{\mu\nu}$ is not invariant under the coordinate transformations $x^\mu \mapsto \tilde x^\mu (x^\nu)$. In fact, as we have discussed in the introduction, this can be considered as the defining property of massive gravity --- a generic massive gravity theory is a theory that breaks some subset of the diffeomorphism invariance of the Einstein--Hilbert gravity. In order to reveal the symmetry breaking pattern of a given massive gravity theory it is useful to restore the diffeomorphism invariance by introducing four St\"uckelberg scalar fields $\phi^A$, which represent the physical coordinates.\footnote{Also for various holographic applications (e.g. the holographic renormalization~\cite{deHaro:2000xn}) it is desirable to have a covariant gravitational theory.} One can then write the mass term using two ingredients: a gauge invariant version of the space-time metric
\be
\I^{AB}\equiv g^{\mu\nu}\pt_\mu\phi^A\pt_\nu\phi^B \; ,
\ee
and a reference metric in the configuration space of the scalar fields
\begin{equation}
f_{AB}(\phi^C) \; ,
\end{equation}
which \emph{can} be a function of the St\"uckelberg fields themselves. The metric $f_{AB}$ is used to raise and lower the scalar fields space indices.\footnote{The reference metric $f_{AB}$ is usually taken to be the flat Minkowski metric $\eta_{AB}$ or is set to coincide with the background metric of the spacetime.} Any mass term written in terms of $\I^{AB}$ and $f_{AB}$ will be manifestly invariant under the general coordinate transformations. In addition, depending on the exact form of the scalar fields Lagrangian, it can be invariant under certain \emph{internal} symmetries of the scalar fields $\phi^A \mapsto \psi^A(\phi^B)$. The field configuration
\be\label{sc0}
\left\langle\phi^A\right\rangle = x^\mu\delta^A_\mu
\ee
spontaneously breaks the internal symmetries and the spacetime diffeomorphisms to a diagonal subgroup --- it is left invariant only by a combination of simultaneous internal field redefinitions and spacetime diffeomorphisms. For example, under the transformations
\be
\phi^A\mapsto\phi^A-\chi^A \; ,\qquad x^\mu\mapsto x^\mu+\xi^\mu
\ee
the background transforms as $\left\langle\phi^A\right\rangle \mapsto \left\langle\phi^A\right\rangle + \xi^A - \chi^A$ and it is left invariant if $\chi^A=\xi^\mu\delta^A_\mu$. This is the diagonal subgroup of the internal and spacetime symmetries.

In general, the scalar fields Lagrangian does not admit the full reparameterisation invariance of the scalar fields due to the fact that the reference metric $f_{AB}$ is fixed and, hence, non-dynamic. Thus, the scalar fields background spontaneously breaks the spacetime diffeomorphisms to the product of the \emph{residual} internal symmetries and their spacetime counterparts. For example, if the reference metric does not depend on the scalar fields, the Lagrangian is invariant under constant shifts $\phi^A\mapsto \phi^A+c^A$ and the diagonal subgroup contains the spacetime translations. An important case is when all the internal scalar field indices are contracted in the Lagrangian. Such theories are invariant under the reparameterisations of the scalar fields that are isometries of the background metric $f_{AB}$. This is the case in the Lorentz invariant massive gravity theories where the Minkowski reference metric leads to the invariance under internal Poincar\'e transformations. In the context of effective field theories of solids and fluids (see~\cite{Nicolis:2013lma,Nicolis:2015sra} and references therein), the Lagrangian is invariant under rigid rotations and volume preserving diffeomorphisms of the scalar fields, respectively~\cite{Dubovsky:2011sj}. The background configuration \eqref{sc0} breaks these symmetries down to the diagonal subgroup of internal and spacetime translations and rotations, or volume preserving diffeomorphisms. In the present work we are making the first steps towards using the effective theories of solids and fluids in the holographic context and applying these theories to the condensed matter systems. 

Leaving some of the scalar field indices not contracted (e.g. when the Lagrangian may contain a sole $\I^{AB}$ as opposed to $\I^{AB}f_{AB}$) breaks the internal scalar field metric isometries to some residual subgroup. In general, this leads to Lorentz violating massive gravity theories which are of the main interest of this paper. In particular, we choose to preserve the internal $O(2)$ rotations of the transverse scalar fields $\phi^I$. This allows us to use $\I^{AB}$ as the building block of our effective Lagrangian with the only condition that the transverse indices have to be contracted with a reference metric proportional to $\delta_{IJ}$. As described above, the full spacetime diffeomorphisms are preserved by the action and are broken spontaneously on the background configuration \eqref{sc0}.

\subsection{Probe limit Lagrangian with $T_{\mu\nu}=0$}\label{sec:probe}
A straightforward but not unique way of constructing a gauge invariant form of the most generic quadratic mass term \eqref{acth} is to set the scalar fields reference metric to be equal to the background black brane metric $\hat g_{\mu\nu}$:
\be
f^{AB}(\phi^C)=\hat g^{\mu\nu}(\phi^C)\delta_\mu^A\delta_\nu^B \;.
\ee
We can then introduce diffeomorphism invariant metric perturbations as
\be\label{hab}
H^{AB}\equiv \I^{AB}-f^{AB}(\phi^r)= g^{\mu\nu}\pt_\mu\phi^A\pt_\nu\phi^B-f^{AB}(\phi^r) \;.
\ee
In the unitary gauge, when the scalar fields are set to be equal to their background solution~\eqref{sc0}, the field $H^{AB}$ equals to the inverse metric perturbations:
\be
H^{AB} = h^{\mu\nu}\delta^A_\mu\delta^B_\nu\,\equiv -\hat g^{\alpha\mu}\hat g^{\beta\nu}h_{\alpha\beta}\delta_\mu^A\delta_\nu^B+\mathcal O\left(h^2\right) \;.
\ee
The covariant form of the action \eqref{acth} is then obtained by replacing the different components of the metric perturbations $h^{\mu\nu}$ with the gauge invariant fields $H^{AB}$ and the explicit $r$-dependence of the masses with an explicit dependence on the radial St\"uckelberg field $\phi^r$. The resulting covariant mass term $\mathcal L_\phi(H^{AB},\phi^r)$ gives a stress-energy tensor that vanishes on the background and, thus, does not contribute to the background equations for the metric. Although the exact vanishing of the backreaction on the metric is a very particular case it can be considered as a probe limit approximation for a generic massive gravity theory. In particular, the number of propagating degrees of freedom as well as the stability requirements of the theory should not be altered in this limit. The exact form of the covariant mass term is given in eqn.~\eqref{actH} in appendix~\ref{sec:linear}, where also the linear stability analysis of the theory is performed. The phenomenological consequences of the mass term \eqref{acth} relevant for the electric response of the dual field theory will be discussed in section~\ref{sec:response}. 

We note here the obvious fact that the number of scalar fields used in our Lagrangian is a matter of choice. However, it limits the available components of $H^{AB}$ and, consequently, the subset of mass terms out of the ones presented in \eqref{acth} that arise if not all four scalar fields $\phi^A$ with $A=t,r,x,y$ are used. Of particular interest for us is the case of only two scalar fields, i.e. $\phi^I$ with $I=x,y$, that is commonly used in the context of holographic massive gravity and will be discussed in great detail in section~\ref{sec:two_fields}. In this case, only the $H^{IJ}$ components can be constructed which implies that all the quadratic mass terms in \eqref{acth} involving $\{t,r\}$ components are absent:
\begin{equation}
\begin{split}
&m_0 =0 \;, \quad m_1=0\;,\quad m_4 = 0 \;, \quad m_5 = 0 \;, \quad m_6 = 0  \;, \quad m_7=0\;,  \\
m_8&=0\;,\quad m_9 = 0 \;, \quad m_{10} = 0 \;, \quad m_{11} = 0 \;,  \quad m_{12} = 0 \;, \quad m_{13} = 0 \;.
\end{split}
\end{equation}
Moreover, since the black brane reference metric $f^{AB}(\phi^r)$ explicitly depends on the absent field $\phi^r$ then even $H^{IJ}$ can only be present in a particular combination that includes its traceless part only, i.e. $(H^{IJ})^2-\frac{1}{2}H^2$. This leads to a quadratic mass term of type \eqref{acth} with
\begin{equation}
m_2^2 = 2 \, m_3^2 \;.
\end{equation}
The same conclusion about the available mass terms could have been reached by noting that the absence of fields $\phi^t,\phi^r$ in the mass term building blocks $\I^{AB}$ and $f_{AB}$ leads to a residual symmetry that is left unbroken by the scalar fields background configuration $\left\langle\phi^I\right\rangle = x^i$. These are the spacetime diffeomorphisms $x^a \mapsto \tilde x^a(x^\mu)$ with $a=t,r$. Hence, all mass terms in \eqref{acth} that transform non-trivially under these diffeomorphisms are forbidden. 

As a result, the above conditions prohibit the appearance of any components of the metric perturbations that are scalars and vectors under the transverse rotations in quadratic action. From the phenomenological point of view it is a case of no particular interest. We note, however, that such strict constraints only arise if we consider generally covariant mass terms with zero background stress-energy tensor, i.e. with zero backreaction on the metric. In practice, the mass Lagrangian $\mathcal L_{\phi}(H^{AB},\phi^r)$ is not the only diffeomorphism invariant mass term that reduces to the quadratic action \eqref{acth} once expanded up to the second order in perturbation theory. The most general covariant mass terms can be obtained by writing the action in terms of the matrix $\mathcal I^{AB}$ and arbitrary functions of the scalar field $\phi^r$. The quadratic perturbative mass term obtained by these other actions will also be of the form \eqref{acth}. However, a general mass Lagrangian of this sort will give rise to a non-zero stress-energy tensor that will backreact on the spacetime metric. As a result, some of the quadratic non-derivative terms arising in the perturbative expansion of $\mathcal L_\phi$ will vanish on the background equation of motion once combined with the Einstein-Maxwell part in the full action \eqref{action}. Hence, there is no one-to-one correspondence of the quadratic non-derivative term of metric perturbations in $\mathcal L_\phi$ and the actual mass terms of the metric perturbations in the case of generally covariant mass potentials with non-zero backreaction.

\subsection{Two fields case: solids and fluids}\label{sec:solids}
A particular example of a massive gravity action with non-zero stress-energy tensor we would like to consider is a generic mass term that depends on only the two transverse St\"uckelberg fields $\phi^I$. Such model is a direct generalisation of the two fields dRGT theory, and provides the simplest way to model momentum dissipation in the dual theory. Since the action does not depend on the $\phi^a$ scalar fields, the background solution $\left\langle \phi^I\right\rangle = x^i\delta^I_i$ leaves the $\{t, r\}$ spacetime diffeomorphisms $x^a \mapsto \tilde x^a(x^\mu)$ unbroken in this case. As before, the most general graviton mass Lagrangian can be written in terms of the two building blocks $\I^{IJ}\equiv g^{\mu\nu} \d_\mu \phi^I \d_\nu \phi^J$ and some reference metric $f_{IJ}$. Since we wish to preserve the internal rotations in the $(I,J)$ plane we set the reference metric to be the identity matrix $f_{IJ}=\delta_{IJ}$. Hence, the mass Lagrangian is an arbitrary function of the powers of $\I$ with all indices contracted with the identity matrix $\delta_{IJ}$. For a $2 \times 2$ matrix there are only two algebraically independent contractions. We choose them to be
\begin{align}
X & \equiv \half \tr[\I^{IJ}] = \half \d_\mu \phi^I \d^\mu \phi^I \; ; \\
Z & \equiv \det [\I^{IJ}] = \half \left( \d_\mu \phi^I \d^\mu \phi^I \d_\nu \phi^J \d^\nu \phi^J - \d_\mu \phi^I \d^\mu \phi^J \d_\nu \phi^I \d^\nu \phi^J \right) \; .
\end{align}
Since any other contraction of $\I$ is a function of $X$ and $Z$, the most general two fields mass term takes the form:
\begin{equation}\label{S2phi}
S_\phi \equiv \int d^4 x \sqrt{-g} \, \mathcal L_\phi=  - \int d^4 x \sqrt{-g} \, V(X, Z) \; .
\end{equation}
The dRGT theory with two fields considered in ref.~\cite{Vegh:2013sk} is a particular case of the two fields massive gravity with the Lagrangian given by
\begin{equation}\label{vdRGT}
V_{\text{dRGT}} = - \beta_1 \sqrt{ \frac12 \left({X + \sqrt{Z}} \right) } - \beta_2 \sqrt{Z} \;.
\end{equation}

The Lagrangian \eqref{S2phi} of the scalar fields is similar to the effective field theories used to describe perfect fluids~\cite{Dubovsky:2011sj} and solids~\cite{Son:2005ak,Endlich:2012pz}. In this context the scalar fields $\phi^I$ denote the comoving Lagrangian coordinates of the material. In order to describe perfect fluids the scalar fields action has to be invariant under field space volume preserving diffeomorphisms
\be
\phi^I\mapsto\psi^I(\phi^J) \; ,\qquad\det\left[\frac{\pt\psi^I}{\pt\phi^J}\right]=1 \; .
\ee
For solids, the action is only invariant under constant shifts and rotations
\be
\phi^I\mapsto\psi^I=O^I_J\phi^J +c^I \; .
\ee
We see that, in general, the scalar fields Lagrangian \eqref{S2phi} describes solids. In the special case when it depends only on the function $Z$, its symmetry is enhanced, and it describes perfect fluids. It is the presence of the field $X$ that ``turns'' the material into a solid. In this sense, the two fields Lorentz violating massive gravity is an effective theory that describes gravity in the presence of a solid or a fluid. We shall present a detailed analysis of the two fields massive gravity in section~\ref{sec:two_fields}.

%
%
%
%
%
%
%

\section{Electric response}\label{sec:response}
In this section we investigate the phenomenological consequences of allowing for the broad class of the graviton mass terms as in \eqref{acth}. We shall concentrate on the dynamics of the vector modes in particular and shall discuss the implications for momentum relaxation and electric response in the dual CFT. Following the discussion of section~\ref{sec:probe} we shall consider the diffeomorphism invariant mass term relevant for vector perturbations 
\be\label{mass}
\mathcal L_\phi = \frac{1}{2}\left(2m_1^2(\phi^r)H^{ti}H^{ti}+m_7^2(\phi^r)H^{ri}H^{ri}+m_8^2(\phi^r)H^{ti}H^{ri}\right) \; .
\ee
For simplicity we assume that the $\phi^r$ dependence of the masses is given by
\be
m_i^2(\phi^r)=M^2(\phi^r)f(\phi^r)^{\alpha_i}m_i^2\;,\\
\ee
where $i=1,7,8$, $f(\phi^r)$ is the emblackening factor written in a gauge invariant form, and $M^2(\phi^r)$ is a universal mass function of dimension mass squared that is regular and non-vanishing at the horizon when $\phi^r=r_h$. The masses $m_i^2$ and the powers $\alpha_i$ are dimensionless constants. A detailed stability analysis of the quadratic Lagrangian in appendices~\ref{sec:linear} and~\ref{sec:stab} leads to conditions $\alpha_8=0$ and $\alpha_1=-\alpha_7=1$. We also find in appendix~\ref{sec:stab} that for arbitrary mass parameters $m_i$ the vector modes of the metric propagate on an effective acoustic background metric that is different from the Reissner-Nordstr\"om background of the Maxwell field. The conditions on the mass parameters for the two effective light cones to coincide read 
\be\label{causal}
m_7^2+2m_1^2=0\;,\qquad\text{and}\qquad m_8^2=0\;.
\ee
A priori these conditions do not need to be imposed as long as the acoustic metric describes a causally stable spacetime~\cite{Babichev:2007dw}. In appendix~\ref{sec:stab} we find no indications that this would not be the case here. Importantly, the mass condition \eqref{causal} is automatically satisfied in the two fields massive gravity described in~\ref{sec:solids}. This means that in these theories both dynamical vector modes propagate on the same effective metric. Since the two fields dRGT theory is a subclass of these theories same conclusion applies. Nevertheless, in what follows we shall leave the mass parameters $m_i^2$ unconstrained unless otherwise specified.

We also note that the mass term \eqref{mass} gives rise to a vanishing stress-energy tensor and as such can be considered as particular. As was already discussed in~\ref{sec:probe} both the stability and phenomenology of mass terms with non-zero backreaction coincide with the results presented below in the probe limit of vanishing graviton mass. 

In appendix~\ref{sec:linear} we find that the dynamics of the vector sector of our model can be described in terms of two gauge invariant vector fields $a_i$ and $\lambda_i$. The first one is the perturbations of the Maxwell field while the second field, $\lambda_i$, was introduced as a Lagrange multiplier and can be expressed in terms of the original fields as
\be
\lambda_i=\frac{\mu}{r_h}a_i-\frac{1}{2r^2}\left(\frac{r^2}{L^2}h_{ti}\right)'+\frac{r^2}{2L^2}\dot h_{ri} \; .
\ee
Henceforth we shall drop the index $i$. The resulting equations of motion for the Fourier modes 
\be\label{fourier}
a(t,r)=a(r)e^{-i\omega t}\;,\qquad \lambda(t,r)=\lambda(r) e^{-i\omega t}\;
\ee
then read
\begin{align}\label{eom1a}
&\left(fa'\right)'+\omega^2\frac{a}{f}-\frac{2\mu^2r^2}{r_h^2}a+\frac{2r^2\mu}{r_h}\lambda=0 \; ,\\\label{eom2a}
&2m_1^2\frac{1}{fM^2}\omega^2\lambda-i\omega m_8^2 \frac{1}{M}\left(\frac{\lambda}{M}\right)'-m_7^2\left(\frac{f}{M^2}\lambda'\right)'+\frac{r^2\det\mathcal P}{2L^2}\left(\frac{\mu}{r_h}a-\lambda\right)=0 \; , 
\end{align}
with $\det\mathcal P=m_8^4-8m_1^2m_7^2\neq 0$. 

\subsection{DC conductivity}\label{sec:dc2}
In this subsection we make use of the analytic approach of ref.~\cite{Blake:2013bqa} for calculating the DC conductivity (for an alternative approach, see~\cite{Donos:2014cya}). The method relies on the existence of a `massless mode' in the bulk that is a linear combination of the Maxwell field $a$ and the graviton, which in our calculations is represented by the field $\lambda$. In zero frequency limit, the existence of this massless mode implies the conservation of a certain quantity $\Pi$ in the radial direction. In~\cite{Blake:2013bqa} it was shown that the quantity $\Pi$ encodes the universal behaviour of the DC conductivity and ensures that it remains constant as one moves from the horizon to the boundary.

Here we shall repeat the analysis for the equations of motion \eqref{eom1a}, \eqref{eom2a} with the mass parameters set to $m_8^2=0,m_7^2=-2m_1^2$ so that to satisfy the condition \eqref{causal}. In this case the equations of motion \eqref{eom1a} and \eqref{eom2a} can be written in the form
\be\label{eomM}
\begin{pmatrix}
L_a&0\\
0&L_\lambda
\end{pmatrix}
\begin{pmatrix} a\\\lambda\end{pmatrix}+\frac{\omega^2}{f}
\begin{pmatrix}a\\\lambda\end{pmatrix}=\mathcal M
\begin{pmatrix}a\\\lambda\end{pmatrix} \; ,
\ee
with the differential operators $L_{a}\equiv f'\pt_r+f\pt_r^2$ and $L_{\lambda}\equiv M^2(f/M^2)'\pt_r+f\pt_r^2$ and the mass matrix $\mathcal M$ defined as
\be
\mathcal M=\begin{pmatrix}
\dfrac{2\mu^2r^2}{r_h^2}&-\dfrac{2\mu r^2}{r_h}\\[1em]
-\dfrac{M^2(r)r^2\det\mathcal P}{4L^2m_1^2}\dfrac{\mu}{r_h}&\dfrac{M^2(r)r^2\det\mathcal P}{4L^2m_1^2}
\end{pmatrix}\;.
\ee
The equations \eqref{eomM} can be compared to the corresponding equations in~\cite{Blake:2013bqa} for the dRGT massive gravity. We find that the dRGT mass 
\be\label{mdRGT}
m^2(r)\equiv-\frac{\beta_1L}{r}-2\beta_2\,
\ee
can be expressed in terms of our parameters $m_1^2$ and $M^2(r)$ as:
\be\label{Mandm}
m^2(r)=2m_1^2r^2M^2(r) \; .
\ee
In particular, the $\beta_1$ term corresponds to the mass function with $\nu=-3$ and the $\beta_2$ term is equivalent to $\nu=-2$.

It is straightforward to check that the mass matrix $\mathcal M$ has a vanishing determinant and, thus, one zero eigenvalue. The eigenvectors $\vec \alpha_1,\vec\alpha_2$ corresponding to the zero and non-zero eigenvalues respectively can be taken to be
\be
\vec\alpha_1=\begin{pmatrix}1\\[1em]\dfrac{\mu}{r_h}\end{pmatrix} \; ,\qquad
\vec\alpha_2=\begin{pmatrix}\dfrac{-\mu L^2}{2r_hM^2(r)m_1^2}\\[1em]1\end{pmatrix} \; .
\ee
The matrix $\mathcal M$ can then be diagonalized as $\mathcal M=UDU^{-1}$ where $D=\text{diag}(\mu_1=0,\mu_2)$ and $U=(\vec\alpha_1\;\vec\alpha_2)$ is a matrix with its column vectors given by the eigenvectors $\vec\alpha_i$. The massless and massive modes, $\delta\lambda_1$ and $\delta\lambda_2$, are then:
\be\label{al_lambdas}
\begin{pmatrix}\delta\lambda_1\\\delta\lambda_2\end{pmatrix}\equiv U^{-1}
\begin{pmatrix}a\\\lambda\end{pmatrix}=\left(\det U\right)^{-1}
\begin{pmatrix}a+\dfrac{\mu L^2}{2r_hM^2(r)m_1^2}\lambda\\[1em]
-\dfrac{\mu}{r_h}a+\lambda\end{pmatrix}\;,
\ee
The equations of motion \eqref{eomM} then take the form
\be
U^{-1}(r)
\begin{pmatrix}
L_a	&	0 \\
0	&	L_\lambda
\end{pmatrix}
\begin{pmatrix}a\\\lambda\end{pmatrix}+\frac{\omega^2}{f}\begin{pmatrix}\delta\lambda_1\\\delta\lambda_2\end{pmatrix}=
\begin{pmatrix}0\\\mu_2\,\delta \lambda_2\end{pmatrix} \; .
\ee
We emphasize that unless the mass function $M^2(r)=\text{const}$ the matrix $U(r)$ is a function of $r$. In the zero frequency limit the massless equation can be written in the form of a radial conservation law
\be\label{momentum}
\Pi'=0 \; ,\qquad \text{with}\qquad\Pi\equiv fa'+\frac{\mu L^2}{2r_hm_1^2}\frac{f}{M^2(r)}\lambda' \; .
\ee
A few remarks are in order. First, we observe that the conserved quantity $\Pi$ can be rewritten in terms of the massless and massive modes as
\be\label{momentum2}
\Pi=f\left(1+\frac{\mu^2L^2}{2r_h^2M^2(r)m_1^2}\right)\delta\lambda_1'+\frac{\mu L^2}{r_hm_1^2}\frac{M'}{M^3}f\delta\lambda_2 \; .
\ee
This form is very similar to the conserved $\Pi$ found in~\cite{Blake:2013bqa} for the dRGT massive gravity. 

Second, we notice that in the unitary gauge the field $\lambda'$ is related to the metric components as
(see equations \eqref{matrix} in appendix~\ref{sec:linear}) relates  to 
\be\label{hlambda}
\lambda' =-2m_1^2M^2(r)\frac{r^2}{fL^4}h_{ti} \; .
\ee
Hence, the conserved quantity $\Pi$, as defined in \eqref{momentum}, reduces to 
\be
\Pi=fa'-\frac{\mu r^2}{r_hL^2}h_{ti} \; .
\ee
This coincides exactly with the radial momentum conjugated to the Maxwell field that can be obtained by varying the on-shell boundary action with respect to the boundary value of the Maxwell field $A_{i(r)}$\cite{Iqbal:2008by,Hartnoll:2009sz}:
\be
\Pi_i=\frac{\delta S_{\text{on-shell}}}{\delta A_{i(r)}}\;.
\ee
The electric conductivity is then defined as the functional derivative
\be\label{ac}
\sigma_i(\omega)=\lim_{r\to 0}\,\frac{1}{i\omega}\frac{\delta\Pi_i}{\delta A_i}=\lim_{r\to 0}\,\frac{1}{i\omega}\frac{a_i'}{a_i} \; .
\ee

The fact that in the zero frequency limit, the canonical momentum with respect to the $r$-foliation of the Maxwell field, $\Pi$, is conserved in the radial direction, was used by Iqbal and Liu in~\cite{Iqbal:2008by} as a motivation to introduce a fictitious membrane DC conductivity for each constant $r$ slice as a response to the massless mode
\be\label{dc_membrane}
\bar\sigma_{DC} (r)=\lim_{\omega\to 0}\,\frac{1}{i\omega}\frac{\delta\Pi(r,\omega)}{\delta (\delta\lambda_1)(r,\omega)} \; .
\ee
We note that this definition of conductivity is equivalent to computing the linear response of the boundary theory to the massless mode $\delta\lambda_1$. In practice, we are interested in the linear response to the Maxwell field perturbations $\delta A$. However, the membrane DC conductivity has nice properties which we would like to exploit in order to extract information about the actual DC conductivity defined as
\be\label{dc_usual}
\sigma_{DC}\equiv\lim_{\omega\to 0}\sigma (\omega)=\lim_{\omega\to 0}\lim_{r\to 0}\frac{1}{i\omega}\frac{a'}{a} \; .
\ee
In the limit $r\to 0$ the two conductivities are related to each other as
\begin{align}
\lim_{r\to 0}\bar\sigma_{DC}&=\lim_{r\to 0}\lim_{\omega\to 0}\frac{1}{i\omega}\left(\frac{\delta A}{\delta(\delta\lambda_1)}\frac{\delta\Pi}{\delta A}+\frac{\delta \lambda}{\delta(\delta\lambda_1)}\frac{\delta\Pi}{\delta \lambda}\right)\nonumber\\
&=\lim_{r\to 0}\lim_{\omega\to 0}\frac{1}{i\omega}\left(\frac{a'}{a}+\frac{\mu^2 L^2}{2r_h^2m_1^2M^2(r)}\frac{\lambda'}{\lambda}\right) \; .
\end{align}
where we have used \eqref{al_lambdas} to express the fields $a,\lambda$ in terms of $\delta \lambda_1,\delta\lambda_2$. 
For the two DC conductivities to coincide we have to demand that the second term in the above expression vanishes near the boundary, i.e. that
\be\label{Lprim}
\lim_{r\to 0}\frac{1}{M^2(r)}\frac{\lambda'}{\lambda}=0 \; .
\ee
In the case of $M^2(r)=\text{const}$ this means imposing the condition $\lambda'(0)=0$ on the boundary. For other choices of mass function $M^2(r)$ one has to similarly find the appropriate condition on the field $\lambda$ at $r= 0$. We shall discuss this in more detail in the next section.

The claim of~\cite{Iqbal:2008by} is that the membrane DC conductivity does not evolve in the radial direction and can be evaluated at arbitrary $r$. In particular, it can be evaluated at the horizon, thus, emphasizing the fact that the DC conductivity of the boundary theory can be entirely determined in terms of the horizon quantities. These ideas were used in the context of massive holography by Blake and Tong in~\cite{Blake:2013bqa}. They have shown that also in the case of massive gravity the membrane DC conductivity \eqref{dc_membrane} is conserved. Thus, the electric DC conductivity on the boundary can be evaluated as the membrane DC conductivity at the horizon. Near the horizon the fields $a$ and $\lambda$ are proportional to $f(r)^{-i\omega/(4\pi T)}$, and the mass function $M^2(r_h)$ is regular. Hence $f\delta\lambda_1' =i\omega\delta\lambda_1$ while $f\delta\lambda_2$ vanishes. Using the equation \eqref{momentum2} we find the DC conductivity:
\be\label{dc_cond}
\sigma_{DC}=\bar\sigma_{DC}(r_h)=1+\frac{\mu^2L^2}{2r_h^2M^2(r_h)m_1^2}=\det U^{-1}(r_h) \; .
\ee
With the identification \eqref{Mandm} this result coincides with the expression for the DC conductivity in the dRGT massive gravity derived in~\cite{Blake:2013bqa}.

\subsubsection*{Phenomenology}
\begin{figure}[h!]
\centering
\includegraphics[width=8cm]{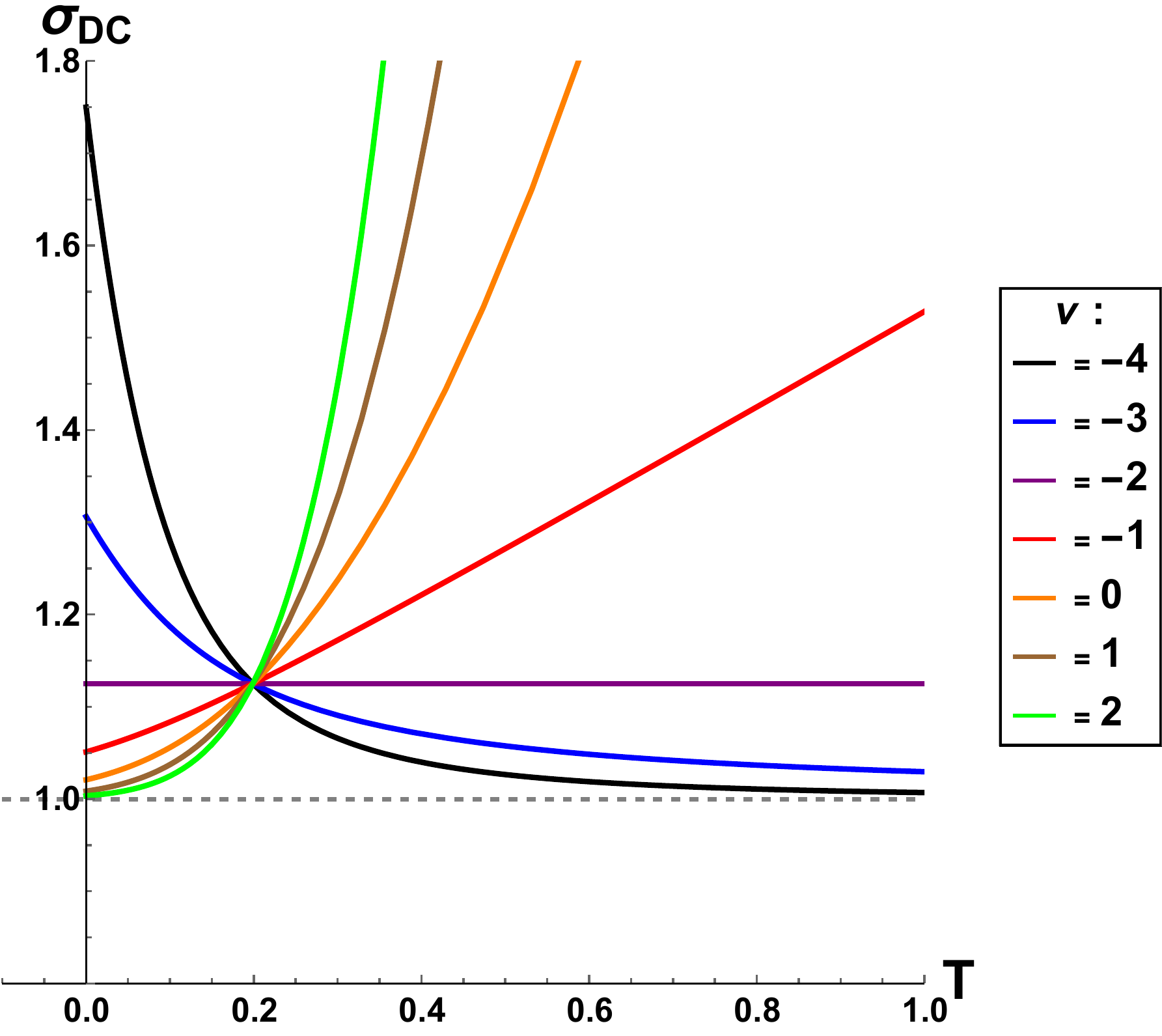}
\caption{The electric DC conductivity (\ref{dc_cond}) as a function of temperature for $\mu=1$ and $m_1=1$ for different values of the parameter $\nu$.}
\label{DCfig}
\end{figure}

A phenomenologically interesting question to investigate is the different types of materials that can be described within the framework of holographic massive gravity and, in  particular, their ability to conduct an electric current. The distinction between different classes of materials is well captured by the temperature dependence of their DC electric conductivity. In the models of massive holography proposed in this paper, it is controlled by the radial dependence of the mass function $M^2(r)$. This determines the dependence of the DC conductivity on the horizon temperature through $T=\left|f'(r_h)\right|/(4\pi)$. For a mass function of the form
\be
M^2(r)=L^{-2}\left(\frac{r}{L}\right)^\nu
\ee
the DC conductivity becomes
\begin{equation}
\sigma_{DC}=1+\frac{\mu^2L^2}{2m_1^2}\left(\frac{L}{r_h}\right)^{2+\nu} \; .
\end{equation}
This defines the distiction between a metallic behaviour ($d\sigma/dT<0$) and an insulating behaviour ($d\sigma/dT>0$). The parameter $\nu$ determines the nature of the dual CFT as shown in fig.~\ref{DCfig}. For $\nu<-2$ the behaviour is metallic while for the opposite case, $\nu>-2$, we are in the presence of an insulator.\footnote{We also note that within holographic models dual to Einstein-Maxwell theory it is impossible to get a properly defined insulator ($\sigma=0$ at $T=0$), as pointed out in~\cite{Grozdanov:2015qia}. Our results suggest that the conclusion reached in~\cite{Grozdanov:2015qia} also holds in field theories dual to massive gravity.} Since the mass function \eqref{mdRGT} in the dRGT theory corresponds to the cases $\nu=\{-2,-3\}$, the dual materials exhibit a metallic behaviour there. The possibility to mimic metallic and insulating behaviour (and a transition between them) in the context of holographic massive gravity has been already pointed out in~\cite{Baggioli:2014roa}. At last, we remark that the special temperature at which all the models with different values of $\nu$ have the same electric conductivity is given by
\be
T_*=\frac{\left|f'(L)\right|}{4\pi}
\ee
and corresponds to the temperature of a black brane with the horizon radius $r_h=L$.

\begin{figure}[h!]
\centering
\includegraphics[width=6.5cm]{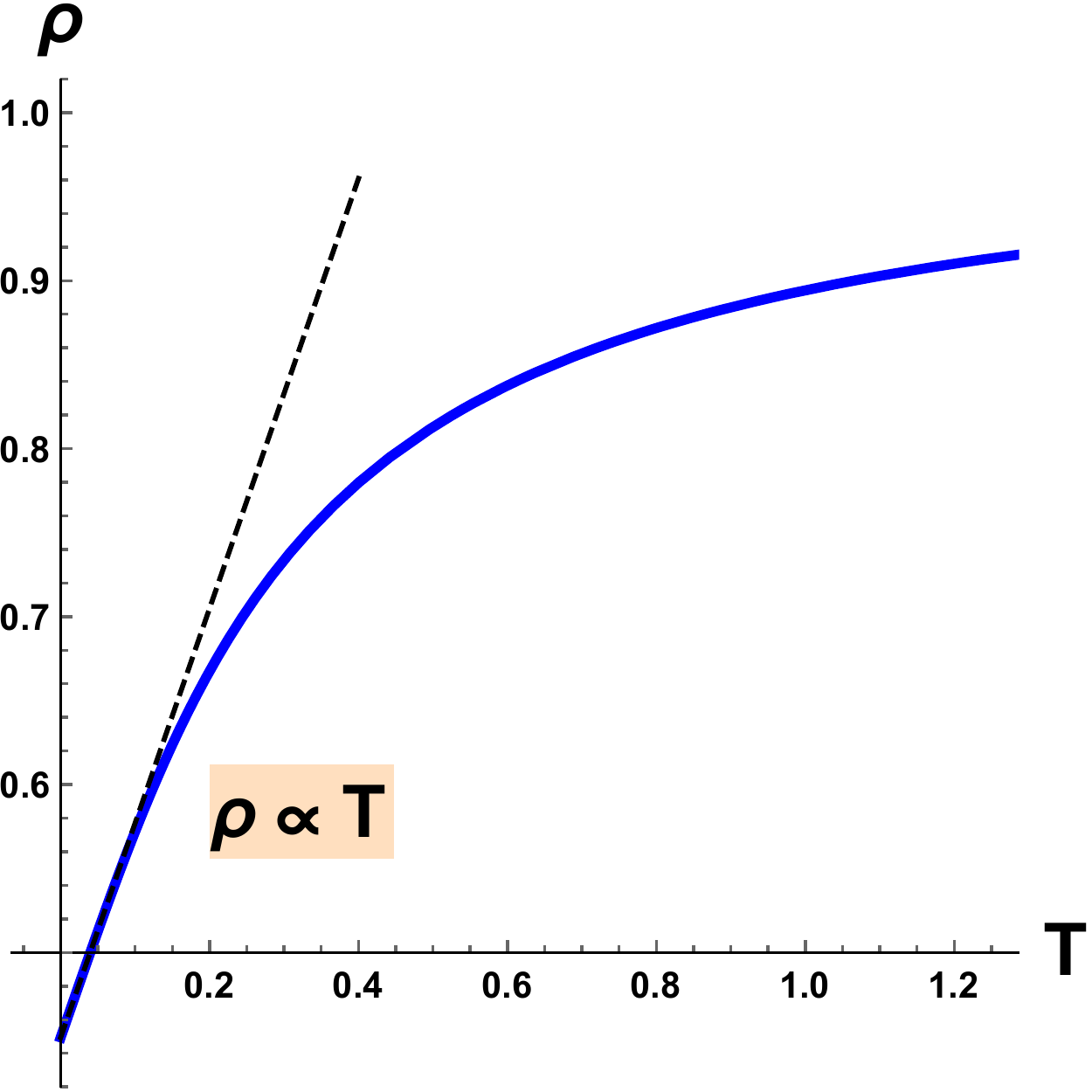}
\caption{The DC resistivity $\rho=1/\sigma_{DC}$ as a function of temperature for parameter values $\nu = -3$, $\mu=1$, $m_1=1$. The dashed line is a linear fit $\rho\propto T$.}
\label{LinTfig}
\end{figure}

We further exploit the freedom offered by the generic power $\nu$ to investigate the possibility of having a linear T resistivity $\rho=1/\sigma_{DC}\propto T^{-1}$. This is a special feature of strange metals (see~\cite{Hartnoll:2009ns} and references therein) which evade the usual scaling predicted by the Fermi liquid theory ($\sigma_{DC}\propto T^{-2}$). Within our class of models we observe a linear scaling in the resistivity at low temperature only for the power $\nu=-3$ as shown in fig.~\ref{LinTfig}. As mentioned above, the $\nu=-3$ mass function coincides with the $\beta_1$ term of dRGT massive gravity. The linear scaling of the DC resistivity for this particular case in the low temperature regime\footnote{At the best of our knowledge, the only holographic example showing linear T resistivity at high temperature is~\cite{Amoretti:2014ola} where a non-trivial dilatonic solution is exploited.} has been already observed earlier in~\cite{Taylor:2014tka}. In~\cite{Davison:2013txa}, the same effect has been seen for a more complicated dilatonic model.

\subsection{AC conductivity}\label{sec:cond}
In order to find the electric conductivity in the dual field theory we need to numerically solve the equations \eqref{eom1a} and \eqref{eom2a} with appropriate boundary conditions on the black brane horizon $r=r_h$ and on the AdS boundary $r=0$. 

\subsubsection*{Boundary conditions on $r=0$}
In order to see what are the appropriate boundary conditions for the field $\lambda$ near the AdS boundary, we expand the equations \eqref{eom1a} and \eqref{eom2a} in the limit $r\to 0$. We use the following ansatz for the radial dependence of the fields:
\be
a=r^\gamma\,,\qquad \lambda=r^\beta\,,\qquad M^2(r)=r^\nu
\ee
and find that the leading order expansion of the fields $a$ and $\lambda$ near the boundary is
\begin{align}
&a=a_0+\frac{r}{L}a_1+\dots\,\\\label{hbound}
&\lambda=\lambda_0+\left(\frac{r}{L}\right)^{\nu+1}\lambda_1+\dots\,.
\end{align}
We see that the first equation coincides with the usual near boundary behaviour of the Maxwell field perturbations in the Einstein-Maxwell theory. The second equation together with \eqref{hlambda} sets the near boundary behaviour of the metric perturbations. For an everywhere constant mass function $M^2=L^2$ we recover the usual result for the metric perturbations $h_{ti}=L^2/r^2\,h_{ti}^{(0)}$ (see e.g.~\cite{Hartnoll:2009sz}). 

For $M^2(r)=r^\nu$ with $\nu+1<0$ the second term in \eqref{hbound} diverges and and the $\lambda_0$ term becomes subleading with respect to the $\lambda_1$ term. The correct near boundary expansion in this case would be 
\be
\lambda = \left(\frac{L}{r}\right)^{|\nu+1|}\left(\lambda_1+\left(\frac{r}{L}\right)^{|\nu+1|}\lambda_0+\dots\right)\;.
\ee
For the regularity of the field $\lambda$ near the AdS boundary we shall thus demand that 
\be\label{Lprim2}
\lambda_1=\lambda'(0)=0\qquad\text{when}\qquad \nu+1<0\,.
\ee 

In section~\ref{sec:dc2} it was shown that the DC conductivity that is finite and conserved in the radial direction is due to the response of the theory to the mode that is massless in the high frequency limit and is given by \eqref{dc_membrane}. In distinction from the Einstein-Maxwell theory the massless mode is a linear combination of both fields, $a$ and $\lambda$. This means that if we wish to consider only the response to the Maxwell field then the additional condition \eqref{Lprim} needs to be imposed. For mass function $M^2(r)=r^\nu$ this reads:
\be\label{Lprim3}
\left.\frac{1}{M^2(r)}\frac{\lambda'}{\lambda}\right|_{r=0}=\left.\frac{\lambda_1}{\lambda_0}\frac{(\nu+1)}{r^{\nu+1}}\right|_{r=0}=0\qquad\rightarrow\qquad \lambda_1=\lambda'(0)=0\quad\text{when}\quad\nu+1>0.
\ee
This ensures that the contribution of the field $\lambda$ to the membrane DC conductivity vanishes. Without this condition, the usual DC conductivity as defined in \eqref{dc_usual} is infinite. In the case when $\nu+1<0$ the expression above automatically vanishes at $r=0$. We notice that the two conditions \eqref{Lprim2} and \eqref{Lprim3} complement each other and impose a general condition that $\lambda'(0)=0$ for all values of $\nu$.

\subsubsection*{Boundary conditions on $r=r_h$}
In order to determine the asymptotic behaviour of the two fields $a$ and $\lambda$ in the vicinity of the horizon we use the ansatz
\be
a=f(r)^\gamma A(r)\,,\qquad \lambda=f(r)^\beta \Lambda(r)\,,
\ee
where the functions $A(r)$ and $\Lambda(r)$ can be expanded as $A(r)=a_0+a_1(r-r_h)+\dots$ and $\Lambda(r)=l_0+l_1(r-r_h)+\dots$. Expanding the equations of motion \eqref{eom1a} and \eqref{eom2a} leads to
\begin{align}\label{horizon}
&(f'(r_h))\gamma)^2+\omega^2=-2r_h\mu\, \frac{l_0}{a_0}f(r)^{\beta-\gamma+1}\,,\\\label{horizon2}
&2m_1^2\omega^2-i\omega m_8^2f'(r_h)\beta-m_7^2(f'(r_h)\beta)^2=-r_h\mu M^2(r_h)\frac{\det\mathcal P}{2L^2}\frac{a_0}{l_0}f(r)^{\gamma-\beta+1}\,.
\end{align}
There are three cases when these equations can be satisfied.

\underline{Case 1:} The first case occurs when $\gamma=\beta$ so that the right hand sides of the above equations become subleading and the equations reduce to
\begin{align}\label{case1a}
&(f'(r_h))\gamma)^2+\omega^2=0\,,\\\label{case1b}
&2m_1^2\omega^2-i\omega m_8^2f'(r_h)\beta-m_7^2(f'(r_h)\beta)^2=0\,.
\end{align}
The first equation gives $\gamma=\pm i\omega/|f'(r_h)|$ and we impose the usual ingoing boundary condition for the Maxwell field $a$:
\be\label{abound}
a(r)=f(r)^{-i\omega/(4\pi T)}A(r)\;,
\ee
where $T=|f'(r_h)|/(4\pi)$ is the black brane temperature. Given the initial assumption $\gamma=\beta$, the equation \eqref{case1b} can only be satisfied when the graviton masses obey the relation
\be\label{Mrel}
m_8^2=-m_7^2-2m_1^2\,.
\ee
This is exactly the condition \eqref{special_masses_0}, discussed in appendices~\ref{sec:vectors} and~\ref{sec:stab}, ensuring that one of the sides of the light cone of the effective acoustic metric $\widetilde g_{ab}$ coincides with one side of the gravitational metric $\hat g_{ab}$ . For this particular choice of the masses we can thus impose the ingoing boundary condition also for the field $\lambda$:
\be\label{lbound}
\lambda(r)=f(r)^{-i\omega/(4\pi T)}\Lambda(r)\;.
\ee

\underline{Case 2:} Another possible choice of the coefficients $\beta$ and $\gamma$ is $\gamma-\beta+1=0$. In this case, the power $\gamma$ is again determined by the equation \eqref{case1a} and, hence, we can use the following boundary conditions for the fields $a$ and $\lambda$:
\be\label{bound_case2}
\lambda(r)=f(r)\cdot f(r)^{-i\omega/(4\pi T)}\Lambda(r)\;,\qquad a(r)=f(r)^{-i\omega/(4\pi T)}A(r)\;.
\ee
The additional dependence on $f(r)$ in the above ansatz for $\lambda(r)$ arises due to the relation $\beta=\gamma+1$. Effectively, such an ansatz results in demanding that the field $\lambda(r)$ vanishes on the horizon. We therefore expect that the effect of the additional degree of freedom introduced by massive gravity is negligible in this case and, thus, cannot help in rendering the DC conductivity finite. Nevertheless, it is interesting to perform a more detailed study of this case due to the fact that, unlike in Case~1, there are no constraints on the mass parameters $m_i^2$. This allows us to probe the parameter space of massive gravity that admits superluminal propagation velocities. 

\underline{Case 3:} The last possibility to satisfy the near horizon equations for integer $\beta$ and $\gamma$ is to set $\beta-\gamma+1=0$. Then the boundary conditions for the two fields read
\be
\lambda(r)=f(r)^{-i\omega/(4\pi T)}\Lambda(r)\;,\qquad a(r)=f(r)\cdot f(r)^{-i\omega/(4\pi T)}A(r)\;.
\ee
However, this amounts to demanding that the Maxwell field vanishes on the horizon which makes this an irrelevant case for the discussion of the electric conductivity and we shall not consider it in the remaining of the paper.

Finally, we note that due to the fact that the mass function $M^2(r)$ is chosen to be regular on the horizon, the particular form of it has no impact on the near horizon boundary conditions.

\subsubsection*{Results and phenomenology}
The AC conductivity can be found by numerically solving equations \eqref{eom1a} and \eqref{eom2a} and using the relation \eqref{ac}:
\be
\sigma(\omega)=\frac{1}{i\omega}\left.\frac{A'(r)}{A(r)}\right|_{r\to 0}\;.
\ee
The numerical results depend on the mass parameters $m_i^2$ and on the functional dependence of the universal mass function $M^2(r)$. Here we consider two different cases depending on whether the relation \eqref{Mrel} is satisfied or not. These correspond to the Case~1 and Case~2 in our above discussion on the near-horizon boundary conditions. 

When the mass relation \eqref{Mrel} is satisfied, we use the ingoing boundary conditions \eqref{abound} and \eqref{lbound} of Case~1 together with the additional condition \eqref{Lprim2}/\eqref{Lprim3} for the derivative of $\lambda$ on the AdS boundary $r=0$. The latter is needed for the finiteness of the DC conductivity and regularity of the solution, as explained above. We also normalize the Maxwell field on the horizon which completes the set of four boundary conditions needed to numerically solve a system of two second order equations for two fields. 

If the mass relation is not satisfied, we use the ingoing boundary conditions \eqref{bound_case2} of Case~2. In this case the near horizon expansion imposes an additional relation between the horizon values of the fields $A(r_h)=a_0$ and $\Lambda(r_h)=l_0$, given in \eqref{horizon2}. It is also necessary to impose the boundary condition \eqref{Lprim2}/\eqref{Lprim3} on the field $\lambda$ at $r=0$. This comprises the set of four boundary conditions needed for solving the system of equations. In distinction from previous case, we cannot anymore normalize the Maxwell field on the horizon since this would overdetermine the system. Nevertheless, the numerical solution for both the Maxwell field and the field $\lambda$ is everywhere regular. However, the DC conductivity has a divergent imaginary part indicating that the real part is infinite. Hence, the values of mass parameters away from the condition \eqref{Mrel} do not provide a well behaved dual description to a field theory with momentum dissipation and we shall not consider this case in what follows.

\begin{figure}
\centering
\includegraphics[width=16.5cm]{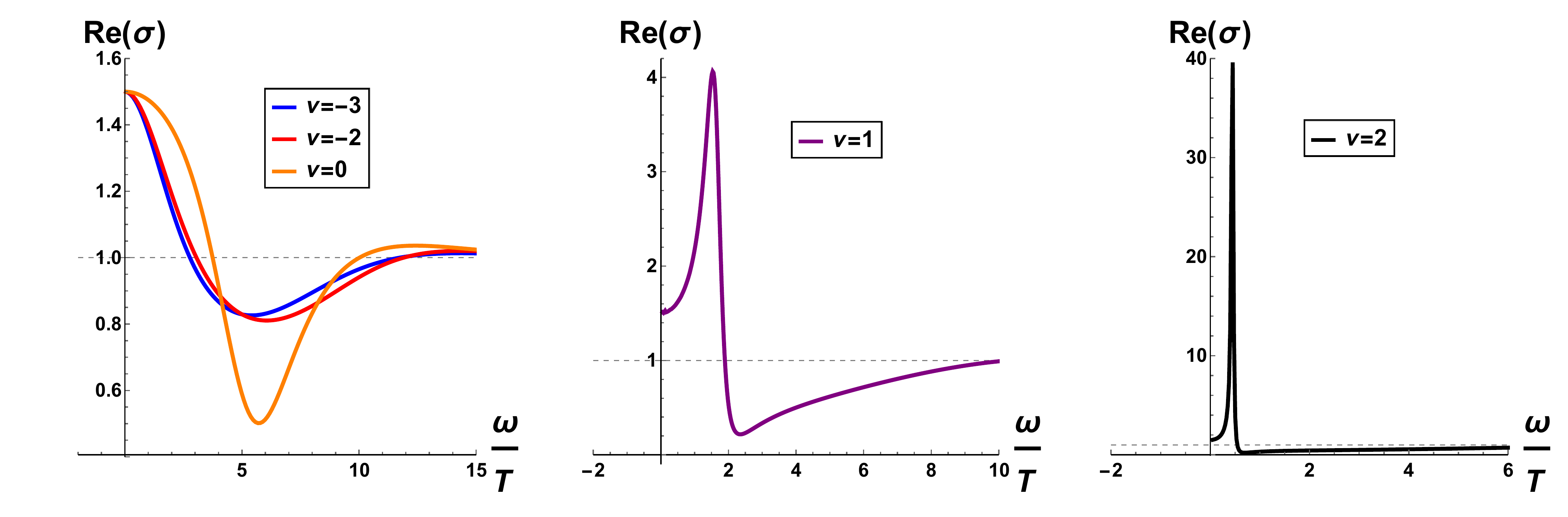}
\caption{Real part of the optical conductivity for different choices of $\nu$ and parameters $\mu=1.5$, $m_1=1$. We fix $r_h=1$ to have the same DC value for different $\nu$.}
\label{ACfig}
\end{figure}

One can solve numerically equations \eqref{eom1a}, \eqref{eom2a} for a generic function $M^2(r)=r^\nu$ and extract the behaviour of the optical conductivity of the system. As already shown in fig.~\ref{DCfig} the power $\nu$ determines whether the model is dual to a metallic or to an insulating state. The choice of $M^2(r)$ influences also the AC response of the dual CFT as shown in fig.~\ref{ACfig}. The value of the real part of the conductivity at zero frequency coincides with the analytic value for the DC conductivity given in \eqref{dc_cond}. For $\nu\gtrsim 0$ we observe a formation of a peak, which becomes sharply localized for higher values of $\nu$ and moves towards zero frequency. This suggests the presence of a localized excitation whose width decreases with $\nu$ very fast. This resonance shows up only in the insulating state and it is reminescent of what has been found in~\cite{Baggioli:2014roa}. It would also be interesting to proceed with an analysis of the quasinormal modes spectrum as in~\cite{Davison:2014lua} to better understand the nature of this pinned response.

We associate the appearance of the peak at these values of $\nu$ with the fact that 
the mass functions that are more localized near the AdS boundary give rise to an explicit source of disorder that seems equivalent to the \emph{lattice disorder} discussed in introduction.
We find this a reasonable interpretation because such a disorder amounts to an additional source of an explicit breaking of translations that makes the otherwise exactly massless phonons become pseudo-Goldstone bosons. 
This is indeed what can be seen in the solid HMGs: one can identify almost massless transverse phonon poles in cases where the breaking is localized near the horizon, while as the profile of the mass function is moved towards the AdS boundary the phonons become massive~\cite{Baggioli:2014roa}.\footnote{Other examples of pseudo-Goldstones that result from a combined spontaneous and explicit breaking of conformal/SUSY/internal symmetries can be found  in refs.~\cite{Megias:2014iwa,Argurio:2014rja}.} Hence, the physical phenomenon related to the appearance of peaks in the electric conductivity for $\nu\gtrsim0$ is expected to be the small collective field excitations --- the phonons --- due to the spontaneous breaking of the translational symmetry. In the presence of charge density this can be interpreted as a polaron formation as first suggested and observed in~\cite{Baggioli:2014roa} (for more on polarons, see~\cite{polaron} and references therein).

\section{Elastic response}
\label{sec:elastic}

In addition to the classification of the materials into solids and fluids according to the residual symmetries that are preserved in the low energy EFT, there is another perhaps more intuitive way to distinguish them, namely, according to the type of response they exhibit under the shape deformation. 
In standard mechanical response theory~\cite{landau7,Lubensky}, the magnitude that encodes the material deformation is a displacement vector $u_i$, and the direct measure of the deformation is the linearized strain tensor,
$$
u_{ij} =\frac{1}{2} \left( \partial_{i} u_{j} + \partial_{j} u_{i} \right)~.
$$
A clear distinction between fluids and solids is that they respond very differently to a \emph{constant} applied shear stress (given by a traceless stress tensor $T_{ij}$ leaving the volume of the material unchanged). For fluids, this leads to a constant shear velocity gradient (traceless part of $\dot u_{ij}$) and the corresponding response parameter is the shear viscosity. For solids, instead, a small constant applied shear stress leads to a constant shear strain (traceless part of $u_{ij}$), and the response parameter is the elasticity. Thus, a practical distinction between solids and fluids is that the static shear elasticity (or rigidity) modulus is nonzero for solids and vanishes for fluids --- fluids do not offer resistance to a constant shear deformation.\footnote{Let us note also that fluids do have a non-zero elastic compression response. Therefore this response does not allow to distinguish between solids an fluids, which is why we shall not consider it here. Let us mention though that the compression response is encoded in HMG in the $m_3(r)$ mass parameter, which is indeed present both for fluids and solids (see section 1). Again, then HMG correctly reproduces that both fluids and solids have an elastic compression modulus.} This distinction between the solid and fluid phases is exactly reproduced in HMG: it is encoded exclusively in the $m_2(r)$ mass parameter, which vanishes for the fluid HMGs but not for the solid HMGs.

Of course, the response of different materials (and also black branes) is more complex than the simplified picture above. Materials can, for instance, exhibit both elastic and viscous ({\em i.e.}, viscoelastic) response. Given that the elastic properties of HMG black branes have not yet been unveiled, we shall restrict here to the elastic response by considering only static applied stress and static deformation or strain. The full viscoelastic response can be studied by considering time dependent applied stresses, but we defer it to a separate publication~\cite{future}.

In homogeneous and isotropic materials, the (static) elastic shear modulus or modulus of rigidity can be defined as the stress/strain ratio 
\be\label{Gdefinition}
T_{ij}^{(T)}=G\; u_{ij}^{(T)} \;,
\ee
where the superscript $T$ stands for the traceless part. 
Equivalently, one can extract the modulus of rigidity, $G$, from a Kubo-like formula that relates it to the Fourier transform at zero frequency and wavenumber of the retarded correlator as:
\be\label{G}
G=\displaystyle\lim_{\omega,k\to0}\text{Re}\vev{T^{(T)}_{xy}\,T^{(T)}_{xy}}^R \; .
\ee
We also note that it differs from the Kubo formula for the shear viscosity given by $\eta = - \displaystyle\lim_{\omega,k\to0}\text{Im}\vev{T^{(T)}_{xy}\,T^{(T)}_{xy}}^R/\omega $, cf.~\cite{Policastro:2001yc}.

\begin{figure}
\centering
\includegraphics[width=16cm]{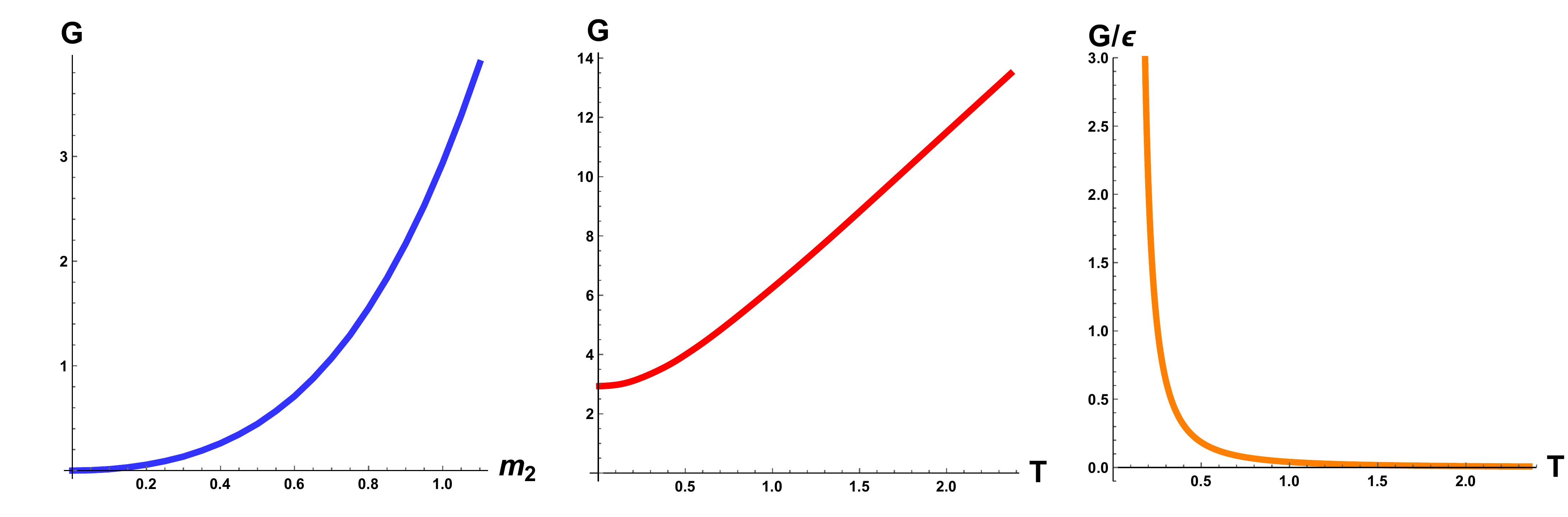}
\caption{Modulus of rigidity $G$ for the mass parameter $m_2^2(r)=r^{-2}$ and chemical potential $\mu=1$: \textbf{Left:} mass dependence; \textbf{Center:} temperature dependence with $m_2=1$; \textbf{Right:} Temperature dependence with $m_2=1$ normalizing the quantity by the T dependent energy density $\epsilon$.}
\label{fig:G(T)}
\end{figure}

Once we have $G$ expressed in terms of the stress tensor two-point functions, it is easy to apply the holographic prescription to extract it. The bulk field dual to $T_{ij}^{(T)}$ is the traceless tensor mode of the metric perturbation 
$h^T_{ij}(r,t,x^k)$. Its equation of motion reads (it can be derived from \eqref{Stensor} in the next section)
\be\label{spin2}
\left[f\partial_r^2 +\left(f'-2\frac{f}{r}\right)\partial_r + \left(\frac{\omega^2}{f}-k^2 -4r^2 m_2^2(r)\right)\right]h_{ij}^T=0 \; .
\ee
For the constant and homogeneous ($\omega=k=0$) response, this equation depends exclusively on the $m_2(r)$ mass parameter. The retarded Greens function is extracted as usual by solving \eqref{spin2} with ingoing boundary conditions and taking the ratio of  the subleading to the leading mode. The resulting response vanishes only for $m_2=0$, i.e. for the fluid HMGs. In fig.~\ref{fig:G(T)} we show the dependence on temperature of the shear modulus for some representative model choice which we take to be $r^2m_2^2(r)=1$. This shows that there is a well-defined sense in which the HMG black branes enjoy a nontrivial  elastic shear response and thus behave as solids. We observe that the rigidity modulus $G$ increases with temperature, whereas most ordinary solids display the opposite dependence and $G$ decreases with increasing $T$. However, in ordinary solids this behaviour occurs at roughly constant energy density while in the middle panel of fig.~\ref{fig:G(T)} the energy density is strongly increasing with $T$ (which relates to the fact that the CFT degrees of freedom are inevitably in a relativistic regime). The ratio of $G$ to the energy density $\epsilon$, instead, does display the more standard decreasing in $T$ form, so in this sense the result seems to be consistent with expectations from ordinary solids. We defer a more complete study of how $G$ depends on other parameters for future work~\cite{future}.

\section{Solid and fluid CFTs}
\label{sec:two_fields}

In this section we study in detail the model with two scalar fields, which provides the simplest effective description of holographic solids and perfect fluids. The two fields mass term \eqref{S2phi} is the most general diffeomorphism invariant Lagrangian that one can write using two scalar fields only. All the stability requirements and holographic predictions can, therefore, be translated directly to the form of the function $V$. This is a great advantage of the two fields massive gravity. However, we shall see below that the phenomenology of \eqref{S2phi} is less rich than that arising from the generic mass terms \eqref{actH} that we consider in appendix~\ref{sec:linear}.

\subsection{Stress-energy tensor and the background solution}

Let us start by asking when such a theory admits an asymptotically AdS black brane solution. In distinction from the mass terms~\eqref{actH}, the scalar fields with action~\eqref{S2phi} provide a non-vanishing contribution to the background stress-energy tensor even on the solution~\eqref{sol}. The scalar fields stress-energy tensor is given by
\begin{equation}
T^\phi_{\mu \nu} = - \frac{2}{\sqrt{-g}} \frac{\delta S_\phi}{\delta g^{\mu \nu}} 
= - g_{\mu \nu} V + \d_\mu \phi^i \d_\nu \phi^i V_X  + 2 \left( \d_\mu \phi^i \d_\nu \phi^i \I^{jj} -  \d_\mu \phi^i \d_\nu \phi^j \I^{ij} \right) V_Z  \; .
\end{equation}
Here and in what follows we drop the distinction between the indices $I$ and $i$, use $V$ with subscripts to denote partial derivatives, e.g., $V_X \equiv \pd V X$, and set $L\equiv 1$. On the solution~\eqref{solmetric},~\eqref{sol} the quantities $X$ and $Z$ take values $\hat X = r^2$ and $\hat Z = r^4$, and the stress-energy tensor takes the form
\begin{align}
\hat T^\phi_{ab} &= - \hat g_{ab} \, \hat V(r) \; , \\
\hat T^\phi_{ij} &= - \hat g_{ij} \left( \hat V(r) - r^2 \, \hat V_X(r) - 2 r^4 \, \hat V_Z(r) \right )\; .
\end{align}
Here $\hat V(r) \equiv V(\hat X, \hat Z)$ is the background value of the Lagrangian. The second line can be rewritten in terms of the function $\hat V (r)$ only as
\begin{equation}
\hat T^\phi_{ij} = - \hat g_{ij} \left ( \hat V(r) - r^2 \frac {d \hat V(r)} {d r^2} \right )\; .
\end{equation}
This structure exactly matches the structure of the Einstein tensor for the black brane metric, i.e. $G^i_i = G^a_a - r^2 \frac d {d r^2} G^a_a$. It thus follows that the two fields massive gravity with the action~\eqref{S2phi} admits an AdS black brane solution with the metric~\eqref{solmetric} for an arbitrary choice of the function $V(X,Z)$. The background solution of the total action \eqref{action} is then completely determined by the $r$-dependence of the background value $\hat V(r)$, with the emblackening factor given by
\begin{equation}\label{f2phi}
f(r) = 1 + r^3 \, \int^r d\tilde r\left(\frac{1}{\tilde r^4} \hat V(\tilde r)+ \frac{\mu^2 \tilde r^4}{2 r_h^2}\right)\; .
\end{equation}
The integration constant parametrises the position of the black brane horizon as well as its mass density and temperature. The fact that the resulting function $f(r)$ is influenced by the graviton mass term only through the background value $\hat V(r)$ means that there are infinitely many different actions that lead to the same background metric. From equation~\eqref{f2phi} it is easy to see that, in principle, it is possible to design a two fields Lagrangian that does not affect the background metric. However, as we show below such theories do not contain any propagating vector modes and cannot reproduce the generic mass term of the form~\eqref{acth}. This is different from the four fields case with zero backreaction studied in appendix~\ref{sec:linear}.

In order to finish the discussion of the background solution we also note that the configuration~\eqref{sol} is a solution to the equation of motion of the scalar fields for any choice of the function $V(X, Z)$. Indeed, the equation of motion has the form:
\begin{equation}\label{eomphi}
\d_\mu \left( \sqrt{-g} \, g^{\mu\nu} \d_\nu \phi^i \, \pd{V}{\mathcal I^{ij}} \right) = 0 \;.
\end{equation}
On the configuration $\phi^i = x^i$, $g_{\mu\nu} = \hat g_{\mu\nu}$ the equation reads
\begin{equation}
\d_i \left( \sqrt{- \hat g} \, \hat g^{ij} \, \pd{\hat V}{\mathcal I^{ij}} \right) = 0 \;,
\end{equation}
and is satisfied since neither the background metric nor the function $\hat V(r)$ depend on $x^i$.

\subsection{Quadratic mass terms}

In order to connect to the generic graviton mass term in \eqref{acth} let us find the quadratic mass terms for the metric perturbations around the black brane. The expansion of the scalar fields Lagrangian up to the second order in inverse metric perturbations in the unitary gauge is given by
\begin{align}\label{V_in_invg}
\mathcal L_\phi  = & - \hat V(r) - \half  \frac{d \hat V}{d r^2} \, h^{ii} 
- \frac18 \frac{d^2 \hat V}{d \left(r^2\right)^2} \, (h^{ii})^2 + \half \hat V_Z \,\left[(h^{ij})^2 - (h^{ii})^2 \right]+ \mathcal O (h^3) \\
 =  & - \hat V(r) + \half r^4 \frac{d \hat V}{d r^2} \, h_{ii} 
- \frac18 \frac{d^2 \hat V}{d \left(r^2\right)^2} \, (h_{ii})^2  - \half r^6 \frac{d \hat V}{d r^2} \, (h_{ij})^2 \notag \\
&
+ \half r^8 \hat V_Z \,\left[(h_{ij})^2 - \half (h_{ii})^2 \right] 
- \half r^4 \frac{d \hat V}{d r^2} \, h_{a i} h_{b j} \, \hat g^{ab}
+ \mathcal O (h^3) \notag
\; ,
\end{align}
where we have used the relations
\begin{equation}
\frac{d \hat V(r)}{d r^2} = \hat V_X(r) + 2 r^2 \, \hat V_Z(r) \;, \qquad  \frac{d^2 \hat V}{\left(d r^2 \right)^2} =  2 \hat V_Z + \hat V_{XX} + 4 r^2 \hat V_{XZ}  + 4 r^4 \hat V_{ZZ}\; .
\end{equation}
It is straightforward to check that the above Lagrangian transforms as a scalar under the $\{t,r\}$ diffeomorphisms which is expected from the fact that the $\phi^a$ fields are absent. However, in situation when the scalar fields provide a non-vanishing background stress-energy tensor, the quadratic expansion of the scalar fields Lagrangian $\mathcal L_\phi$ alone is insufficient to describe the graviton mass term. Instead we shall consider the quadratic expansion of the full action \eqref{action} including the Einstein-Maxwell part. As expected, the vector, tensor, and scalar modes come in the quadratic expansion separately. 

The action for the vector modes takes the same form as the vector mode action~\eqref{act_vec} found in appendix~\ref{sec:vectors} with the condition \eqref{cond_alpha} satisfied and with the mass parameters\footnote{According to these definitions, the dimensionless mass parameters $m_i^2$ take the values $2m_1^2=-m_7^2=1$}
\begin{equation}
2 m_1^2(r) = f(r) M^2(r) \; , \qquad m_7^2(r) = -f(r)^{-1} M^2(r)  \; , \qquad m_8^2 = 0\;.
\end{equation}
The mass function $M^2(r)$ introduced in section~\ref{sec:response} is given in terms of the function $V(X, Z)$:
\begin{equation}\label{m_vec}
M^2(r) = \frac1{r^2} \frac{d \hat V(r)}{d r^2} \; .
\end{equation}
These masses automatically have the correct powers of $f(r)$ and satisfy the conditions~\eqref{causal}, which guarantee that the gravitational vector mode propagates on the light cone defined by the background metric. The mass parameter of the vector modes, the analogue of the $m^2(r)$ in the two fields dRGT theory specified in \eqref{mdRGT}, is given by $r^2 M^2(r) = \frac{d \hat V(r)}{d r^2}$ and it is completely determined by the background value of the scalar fields Lagrangian. The expression~\eqref{dc_cond} for the DC conductivity in the two fields theory takes the form
\begin{equation}
\sigma_{DC} = 1 + \mu^2 \left( \frac{d \hat V(r_h)}{d r^2} \right)^{-1} \; .
\end{equation}
In a generic two fields massive gravity the $r$-dependence of this mass parameter can be more general than in the dRGT case~\eqref{mdRGT} and is constrained only by the stability requirements. In particular, from the quadratic action of the vector sector~\eqref{act_lam_a} one can see that for the mode $\lambda$ to have a healthy kinetic term the vector modes mass parameter has to be positive,
\begin{equation}
\frac{d \hat V(r)}{d r^2} > 0 \;.
\end{equation}

We also note that the functions $V(X,Z)$ with constant background values $\hat V(r)$ are special. In this case the scalar field background contribution to the stress-energy tensor takes the form of a cosmological constant and can be reabsorbed in the definition of $3/L^2$. Therefore, the scalar fields have vanishing stress-energy tensor and do not affect the background metric. However, it also leads to the vanishing of the vector mass parameter~\eqref{m_vec}. In the vector mode analysis of section~\ref{sec:vectors} it corresponds to the case when all the mass parameters relevant for the vector sector are vanishing, $m_1^2(r) = m_7^2(r) = m_8^2(r) = 0$. The vector sector of such theory at quadratic level is identical to the one of pure Einstein-Maxwell theory and does not contain any propagating gravitational degrees of freedom. On the CFT side it implies the absence of momentum dissipation and infinite value of electric DC conductivity. We conclude that the two fields action cannot provide a healthy graviton vector mode without contributing non-trivially to the background stress-energy tensor. It also implies, that in order to have a holographic massive gravity with propagating vectors and finite DC conductivity, which in the meantime does not contribute to the background stress-energy tensor, one has to add more than two scalar fields.

Let us now turn to the helicity-two tensor mode of the metric. In 3+1 dimensional bulk this mode exists only for perturbations homogenous in the transverse directions and encodes the viscoelastic response of the boundary theory in the holographic description. The quadratic action for the traceless tensor mode takes the form
\begin{equation}\label{Stensor}
S = \int d^4 x\, \frac1{4r^2} \left( \frac1{f(r)} (\dot h_{T})^2 - f(r) (h'_{T})^2 - 2 \hat V_X(r^2) \, h_{T}^2\right) \; ,
\end{equation}
where $h_{T}$ stands for any of the two helicity-two components, which we parameterise as $h_+ \equiv \half \frac{L^2}{r^2} \big(h^{xx} - h^{yy} \big)$ and $h_\times \equiv  \frac{L^2}{r^2} h^{xy}$.
This action is equivalent to the transverse traceless sector of the generic quadratic mass term~\eqref{acth} with the mass parameter $m_2^2(r)$ given by
\begin{equation} \label{m_ten}
m^2_2(r) = \frac1{2r^2} \hat V_X(r) \;.
\end{equation}
The mass of the tensor mode~\eqref{m_ten} is not completely determined by the background behaviour of $\hat V(r)$, and is thus independent of the mass of the vector mode and of the value of the DC conductivity. We would like to note, that the tensor mass vanishes for the two field actions that only depend on $Z$, i.e., for the theories describing perfect fluids. This is to be expected, since such theories possess an additional symmetry of transverse space diffeomorphisms that forbids the appearance of the mass term for the helicity two mode. This observation also agrees with an earlier finding in the two fields dRGT massive gravity in~\cite{Alberte:2014bua} that the $\beta_2$ term in the mass term \eqref{vdRGT} gives zero contribution to the mass of the transverse graviton. Hence, in dRGT theory the $\beta_1$ term describes solids, while $\beta_2$ term corresponds to a fluid. 

In accordance to the symmetry considerations of the section~\ref{sec:solids}, the scalar sector of the theory does not contain any dynamical degrees of freedom. In particular, the mass parameters $m_{10}$ and $m_{12}$ of the quadratic action vanish, and as shown in appendix~\ref{sec:scalars} all the scalar degrees of freedom can be eliminated by using constraints and are thus non-dynamical. 

\subsection{Transverse phonons}
In order to analyse further the stability of the two fields theory we proceed with the study of the quadratic action for the inhomogeneous vector modes in the decoupling limit and derive their propagation speeds. For this we consider the quadratic action for the scalar field perturbations $\pi^i = \phi^i - x^i$ while keeping the metric to be fixed, i.e. $g_{\mu\nu} = \hat g_{\mu\nu}$.
%
In the presence of $x^i$ dependence, the two component vector $\pi^i$ can be split in the longitudinal and transverse parts in the following way
\begin{equation}
\pi^i = \frac{\d^i}{\sqrt{-\Delta}} \pi_L + \frac{\epsilon^{ij} \d_j}{\sqrt{-\Delta}} \pi_T \;.
\end{equation}
The quadratic Lagrangian for these components then reads
\begin{align}
\mathcal L^{(2)}_\phi &= - \half r^2 \frac{d \hat V}{d r^2} \, (\d_a \pi_T)^2 - \half r^2 \hat V_X \, (\d_i \pi_T)^2 \notag\\
&- \half r^2 \frac{d \hat V}{d r^2} \, (\d_a \pi_L)^2 - \half r^2 \left( \hat V_X + r^2  \frac{d^2 \hat V}{\left(dr^2 \right)^2} \right) \, (\d_i \pi_L)^2 \;,
\end{align}
where the repeated $a$ indices are contracted with the metric $\hat g^{ab} r^{-2}$. In accordance with the full analysis of the homogeneous vector modes, the absence of ghost requires that $ \frac{d \hat V}{d r^2}$ is positive. We can also read off the propagation speeds of both vector modes:
\begin{align}
c_T^2 &= f(r) \hat V_X \left( \frac{d \hat V}{d r^2} \right)^{-1} = f(r) \frac{2 m_2^2(r)}{M^2(r)} \; , \\
c_L^2 &= f(r) \left( \hat V_X + r^2  \frac{d^2 \hat V}{\left(dr^2 \right)^2} \right ) \left( \frac{d \hat V}{d r^2} \right)^{-1}\; .
\end{align}
The absence of gradient instabilities puts additional constraints on the function $V(X,Z)$
\begin{align}
\hat V_X \ge 0 \; ,\qquad \hat V_X + r^2  \frac{d^2 \hat V}{\left(dr^2 \right)^2} \geq 0 \; .
\end{align}
One notices that the speed of propagation of the transverse phonons $c_T$ is proportional to the $m_2$ mass parameter, which is the one that is forced to vanish by the residual VPDiffs in the fluid case.  
Hence, exactly as it happens for fluids in flat space, the VPDiffs are responsible for decoupling the transverse phonons, $\pi^T_i$, by forcing them to be non-propagating (the dynamics of these degrees of freedom is related to  vortices in a quite interesting way, see~\cite{Endlich:2010hf}). 
That the vanishing of $c_T$ is equivalent to the vanishing of the tensor mode mass $m_2$ can be easily understood in terms of the St\"uckelberg trick. The reason is that, in the space part of the metric, $h_{ij}$ is accompanied by $\partial_{(i}^{}\pi_{j)}^T$ to form a gauge invariant quantity, and so the mass term for the transverse traceless mode and the spatial gradient term for the transverse phonons $\pi^T_i$ are in fact the same thing.\footnote{See~\cite{Endlich:2012pz} for how this works in the cosmological solutions produced by solids/fluids. In the fluid limit, $c_T\to0$ and the tensor modes are massless.} Notice, however, that the fact that the tensor mode mass must vanish for fluids is not completely obvious from the nonlinear form of the metric potential \eqref{Vsf}, which is a function of $\det g^{ij}$. Naively, it seems that it can still give rise to a mass term for tensor modes. However, this cannot happen and the simplest physical reason is that the transverse phonons in a fluid must have vanishing spatial gradient term. In turn, this further implies that black brane solutions acquire rigid properties in solid HMGs but not in fluid HMGs.

\subsection{Linear and nonlinear consistency}
Let us stress that the two fields massive gravity is free from the Boulware-Deser ghost~\cite{Boulware:1973my}, i.e. that at full nonlinear level the number of unconstrained degrees of freedom is the same as at the linear level. Showing this is almost trivial if one does not go to the unitary gauge, where $\phi^i$ are frozen to their vev $\vev{\phi^i}=x^i$, but considers it as a theory of two scalar fields minimally coupled to gravity, cf.~\cite{Alberte:2013sma}. For the most general Lagrangian that depends on the first derivatives of the scalar fields, $V(X, Z)$, the equation of motion of the scalar fields is given in \eqref{eomphi} and schematically it looks like 
\begin{equation*}
\pd{V}{\mathcal I^{ij}} \Box\phi^i +  \frac{\partial^2V}{\partial \mathcal I^{ij} \partial \mathcal I^{kl}}\d^\mu \phi^i \d^\nu \phi^k \; \nabla_\nu \partial_\mu  \phi^l  = 0 \;.
\end{equation*}
Since this equation is of the second order in time and space derivatives even at full nonlinear level, the full dynamics of the model is completely determined by specifying the same initial data as for the  linearized problem. This means that there are no additional degrees of freedom showing up at nonlinear level. The other dynamical equations of the problem, the Einstein equations, do not alter this conclusion,  because the stress tensor from $\phi^i$ is of first order in derivatives. It is crucial that the scalar fields $\phi^i$ are minimally coupled to gravity, so that no additional degrees of freedom could arise in the gravity sector.

Note that the same conclusion is not so easy to reach if one starts directly in the unitary gauge. In that gauge, the two fields theory looks like adding a potential term for the metric $V(\tr(g^{ij})\, , \det(g^{ij}) )$ and the only dynamical equations are the Einstein equations. Nevertheless, because of the above argument, this theory is completely free from the BD problem {\em for arbitrary choice of $V(X,Z) $}. The theory propagates four degrees of freedom at non linear level in total. There are separate conditions to be imposed in order to insure that the two additional degrees of freedom are healthy. But this is different from the BD ghost.

\section{Conclusions}\label{sec:conclusions}
In this paper we have analyzed theoretical consistency and phenomenological implications of a wide class of holographic massive gravity theories. These encompass all possible Lorentz violating graviton mass terms that are compatible with symmetries of the AdS black brane background solution and preserve the homogeneity and isotropy in the two spatial directions of the dual field theory. 

Such theories are known to provide a holographic framework for momentum relaxation and are used as a phenomenologically viable description of transport properties in various condensed matter systems. However, only very particular massive gravity model, i.e. the dRGT theory, has been mostly exploited for this purpose so far. Such restriction is justified in the Lorentz invariant massive gravity where it is well-known that a generic massive gravity is plagued by a ghost instability. In the Lorentz violating case, however, there are many more graviton mass terms that are allowed by theoretical consistency. For the first time, in this work we have spelled out all the mass terms that are permitted by the symmetry of the black brane metric in the spirit of a similar analysis that has been carried out earlier for the Minkowski background in~\cite{Rubakov:2004eb,Dubovsky:2004sg}. We have shown how generally covariant form of these masses can be reconstructed by the usual St\"uckelberg trick by using a set of four scalar fields. 

A large part of the paper is devoted to the discussion of massive gravity theories where only two transverse St\"uckelberg fields are employed. We argue that such theories are the covariant version of the effective field theories that describe solids and fluids. As such, also the two fields holographic massive gravity can be broadly split into theories describing solids and fluids. Due to the enhanced symmetry in fluids, the set of allowed mass terms is more constrained in this case. Of particular importance for this work is the observation that the $m_2$ mass term is absent for fluids. We further show that this mass term implies important phenomenological consequences for the response in the dual field theory. We find that theories with $m_2$ mass term enjoy a non-zero response to static shear deformations thus signifying a sharp distinction between holographic solids and fluids. 

We have also investigated the phenomenological consequences of the new classes of the holographic massive gravities on the electric response in the dual theory. We find that it can be parametrised by a universal mass function $M^2(r)$ that encodes the electric properties of the dual system. We show that this function is proportional to the $m^2(r)$ mass parameter in the dRGT theory. However, in our case the form of this function is free. We generalize the universal expression for the DC conductivity found for the dRGT massive gravity in~\cite{Blake:2013bqa} for this case. By assuming the particular dependence $M^2(r)\propto r^\nu$ we find that the parameter $\nu$ determines whether the dual material has the properties of an insulator or a metal. It also regulates the appearance of the phonons in the dual theory. Moreover, we also show that for $\nu=-3$ the DC resistivity at low temperatures scales linear with temperature. The same phenomenon has been previously observed in the dRGT theory in~\cite{Taylor:2014tka}. Let us add here that the metric potential $V(X,Z)$ that gives rise to this behaviour scales like $V\sim \sqrt X$ (or equivalently $V\sim Z^{1/4}$) which is similar to the square root structure in the scalar DBI Lagrangian. Since the shape of the latter is protected by a specific reparametrization-like symmetry it might be that also the universal linear in $T$ resistivity behaviour follows from a symmetry principle.

The case of holographic massive gravity with all four St\"uckelberg fields employed has never been considered elsewhere. Here we perform a thorough analysis of the homogeneous scalar, vector and tensor perturbations of these theories. We find that the problematic scalar sector can be healthy and propagate up to one helicity-0 degree of freedom if the mass parameters, as defined in \eqref{acth}, satisfy
\be
4m_0^2m_{10}^2-m_{12}^4=0\;.
\ee
The vector sector is in general stable and contains two dynamical vector fields with two degrees of freedom each. However, we find that the effective light cones on which these two fields propagate are, in general different from each other. While the Maxwell field perturbations propagate on the background Reissner-Nordstr\"om geometry, the second vector field, $\lambda$ propagates on an effective geometry determined by the graviton mass parameters. The two light cones partially coincide only in the case when the mass parameters satisfy
\be
m_8^2+m_7^2+2m_1^2=0\;.
\ee
The light cone of the gravitational vector mode, although apparently causally stable, can be wider than that of the Maxwell field and thus allows superluminal propagation velocities. Although bearing no apparent pathologies we find that the superluminal sector of the holographic massive gravities is not suitable for describing momentum relaxation in holographic framework. The reason for this is that the ingoing boundary conditions for the field $\lambda$ cannot be imposed in the usual manner in this case. In the theories, where the light cones of both vector modes coincide, the electric response of the dual field theory can also be parametrized by a single mass function $M^2(r)$ and is identical to the two fields case.

\section*{Acknowledgments}
We thank Alessandro Braggio, Nicodemo Magnoli, Daniele Musso, Alex Pomarol and Sebastien Renaux-Petel  for very useful discussions. MB would like to thank University of Illinois, ICMT and Philip Phillips for the warm hospitality during the completion of this work.
We acknowledge support by the Spanish Ministry MEC under grants FPA2014-55613-P and FPA2011-25948, by the Generalitat de Catalunya grant 2014-SGR-1450 and by the Severo Ochoa excellence program of MINECO (grant SO-2012-0234).

\appendix

\section{Second order action - Linear consistency}\label{sec:linear}
In this section we shall study the linear stability of the generally covariant graviton mass term with four St\"uckelberg fields described in section~\ref{sec:probe} and explicitly given by:
    \begin{align}\label{actH}
\mathcal L_\phi (H^{AB},\phi^r)&= \frac{1}{2}\left ( m_{0}^2(\phi^r) (H^{t t})^2+ 2m_{1}^2(\phi^r) H^{t i} H^{t i } - m_2^2(\phi^r) H^{i j} H^{i j}  \right. \notag \\
&\quad  + m_{3}^2(\phi^r) H^{i i} H^{j j} -2m_{4}^2(\phi^r) H^{t t} H^{i i} \notag\\
&\quad  + m_{5}^2(\phi^r) (H^{r r})^2 + m_{6}^2(\phi^r) H^{t t} H^{r r }    \\
&\quad   +m_{7}^2(\phi^r) H^{r i} H^{r i } + m_{8}^2(\phi^r) H^{t i} H^{r i } +m_{9}^2(\phi^r) H^{r r} H^{i i} \notag\\
&\quad + m_{10}^2(\phi^r)H^{rt}H^{rt} +m_{11}^2(\phi^r)H^{rt}H^{ii}+m_{12}^2(\phi^r)H^{tt}H^{rt}+m_{13}^2(\phi^r)H^{rr}H^{rt}\big ) \;.\notag
\end{align}
For the analysis of quadratic Lagrangian it is enough to expand the gauge invariant fields $H^{AB}$ defined in \eqref{hab} up to first order in perturbations:
  \be
  H^{AB}=h^{\mu\nu}\delta_\mu^A\delta_\nu^B+\hat g^{\mu\nu}\delta^A_\mu\pt^\nu\pi^B+\hat g^{\mu\nu}\delta^B_\mu\pt^\nu\pi^A-\pt_r\hat g^{AB}\pi^r\;,
  \ee
  where the last term arises from the perturbation of the reference metric $f^{AB}$. We shall classify the perturbations according to the scalar, vector, and tensor representations of the transverse $O(2)$ rotation group as in \eqref{scalar}-\eqref{tensor}. At quadratic level the Lagrangians of the three sectors decouple from each other and will therefore be studied separately.
 
\subsection{Vector modes}\label{sec:vectors}
We consider the mass term relevant for vector perturbations 
\be
\mathcal L_\phi = \frac{1}{2}\left(2m_1^2(\phi^r)H^{ti}H^{ti}+m_7^2(\phi^r)H^{ri}H^{ri}+m_8^2(\phi^r)H^{ti}H^{ri}\right)\,
\ee
and assume for simplicity that the $\phi^r$ dependence of the masses is given by
\be\label{mansatz}
m_i^2(\phi^r)=M^2(\phi^r)f(\phi^r)^{\alpha_i}m_i^2\;.\\
\ee
Here $i=1,7,8$, $f(\phi^r)$ is the emblackening factor written in a gauge invariant form, and $M^2(\phi^r)$ is a universal mass function that is regular and non-vanishing at the horizon $\phi^r=r_h$. For concreteness, we also assume that the mass function $M^2(\phi^r)\geq 0$ everywhere and has the dimension of mass squared. The masses $m_i^2$ on the right hand side of \eqref{mansatz} and the powers $\alpha_i$ are dimensionles constants. 

As was clarified in section~\ref{sec:setup}, vector perturbations are described by the components
\be
h_{ti}\equiv \frac{L^2}{r^2}\tilde h_{ti},\quad h_{ri}\equiv\frac{L^2}{r^2}\tilde h_{ri},\quad a_i,\quad\pi^i  \; .
\ee
Since the mass term above is diffeomorphism invariant by construction it is possible to rewrite it in terms of gauge invariant fields. Indeed, we find that up to the first order in perturbations the fields $H^{ir}$ and $H^{it}$ can be written as
\be
H^{ir}=\hat g^{rr}\bar{\pi}^i_r \; ,\qquad H^{it}=\hat g^{tt}\bar{\pi}^i_t \; ,
\ee
where $\bar{\pi}^i_r$ and $\bar{\pi}^i_t$ are the gauge invariant fields introduced in~\cite{Alberte:2014bua} and are defined as
 \begin{align}\label{vec}
 &\bar\pi^i_r\equiv(\pi^i)'-\tilde h_{ri} \; ,\\
 &\bar\pi^i_t\equiv\dot\pi^i-\tilde h_{ti}\;.
 \end{align}
 The Maxwell perturbation $a_i$ is diffeomorphism invariant by itself, and there is one more useful gauge invariant quantity
 \be\label{constraint}
  \bar h_i\equiv \tilde h_{ti}'-\dot{\tilde h}_{ri} \; ,\qquad \bar h_i =\dot{\bar\pi}^i_r-(\bar\pi^i_t)'\;
  \ee
 that is, however, related to the other variables. With these definitions at hand, it is straightforward to rewrite the total action \eqref{action} up to the second order in perturbations in a diffeomorphism invariant form
\begin{align}\label{act_vec}
\sqrt{-g}\,\L=\frac{L^2}{2r^2f}
&\left(- \frac{2\mu r^2}{r_h}fa_i\bar h_i+\frac{1}{2}f\bar h_i^2-r^2f^2 a_i'^2+r^2\dot{a}_i^2\right)\notag\\
+\frac{M^2(r)}{2f^2}&\bigg(2m_1^2f^{\alpha_1}(\bar{\pi}^i_t)^2+m_7^2f^{4+\alpha_7}(\bar{\pi}^i_r)^2-m_8^2f^{2+\alpha_8}\bar{\pi}^i_t\bar{\pi}^i_r\bigg) \;.
\end{align}
The above action depends on the Maxwell field $a_i$ and the three fields $\bar h_i,\bar{\pi}^i_t,\bar{\pi}^i_r$ that are related to each other via the constraint \eqref{constraint}. The vector indices of the fields are always contracted trivially, and in what follows we shall suppress them in order to simplify the appearance of the expressions.

Since one expects at most two dynamical vector fields in the theory -- the Maxwell field and the helicity-one part of the massive graviton -- it should be possible to integrate out two of the four vector fields. In order to consistently impose the constraint \eqref{constraint} we add it to the Lagrangian with a Lagrange multiplier $\lambda$:
\begin{equation}\label{Lmult1}
\sqrt{-g}\,\L_\lambda = \lambda L^2\left( \bar h - \dot{\bar\pi}_r + (\bar\pi_t)'  \right) \;.
\end{equation}
The action \eqref{act_vec} together with the above constraint contains five vector fields $\bar\pi_t,\,\bar\pi_r,\,\bar h,\,a$, and $\lambda$, all of which should be treated independently. The corresponding equations of motion read 
\begin{align}\label{matrix}
&M^2(r)\begin{pmatrix}
-m_8^2f^{\alpha_8}&2m_7^2f^{2+\alpha_7}\\
4m_1^2f^{\alpha_1-2} &-m_8^2f^{\alpha_8}
\end{pmatrix}
\begin{pmatrix}
\bar{\pi}_t\\
\bar{\pi}_r
\end{pmatrix}\equiv M^2(r)\mathcal P\begin{pmatrix}
\bar{\pi}_t\\
\bar{\pi}_r
\end{pmatrix}=2L^2
\begin{pmatrix}
-\dot\lambda\\
\lambda'
\end{pmatrix}\\\label{hfield}
&\quad\bar h = \frac{2\mu r^2} {r_h} \, a- 2 r^2 \lambda \; , \\
&\quad\bar h = \dot{\bar\pi}_r - (\bar\pi_t)' \; , \\
&\quad\ddot{a} - f^2 a''-ff'a'+\frac{\mu f}{r_h}\bar h = 0 \;. 
\end{align}
The equations \eqref{matrix} and \eqref{hfield} can be used to eliminate the fields $\bar{\pi}_t,\bar{\pi}_r$ and $\bar h$ from the action \eqref{act_vec} and rewrite it in terms of $a$ and $\lambda$. We notice that the system of equations \eqref{matrix} has a non-degenerate solution only in the case when $M^2(r)\neq 0$ and 
\be
\det\mathcal P=m_8^4f^{2\alpha_8}-8m_1^2m_7^2f^{\alpha_1+\alpha_7}\neq 0 \; .
\ee
In order to satisfy the above condition at the horizon for a generic choice of the masses $m_i^2$ one has to impose
\be\label{cond_alpha}
\alpha_8=0\;,\qquad \alpha_1=-\alpha_7\equiv\alpha\;.
\ee
Under these assumptions the $\det\mathcal P$ is constant and non-vanishing everywhere. The case when $\det\mathcal P=0$ marks one particular phase of the Lorentz violating massive gravity and has to be considered separately. 
 
 When $\det\mathcal P\neq 0$ we can solve for the fields $\bar{\pi}_t,\bar{\pi}_r$ and $\bar h$ algebraically. After substituting the solutions back in the Lagrangian \eqref{act_vec} with the constraint term added we obtain:
 \begin{align}\label{act_lam_a}
\sqrt{-g}\,\L = L^2 &\left ( \frac1{2 f} \, \dot a^2 - \frac f2 \, (a')^2 
- r^2 \left( \lambda - \frac\mu{ r_h} \, a \right)^2\right)\notag\\
+&\frac{2L^4}{M^2(r)\det\mathcal P}\bigg(2m_1^2f^{\alpha-2}\dot\lambda^2+m_7^2f^{2-\alpha}\lambda'\,^2-m_8^2\dot\lambda\lambda'
\bigg) \;.
\end{align}

When $\det\mathcal P=0$ we find that the Maxwell field $a$ and the field $\lambda$ decouple from each other. Moreover, $\lambda=0$ by its equation of motion and the Lagrangian for $a$ reduces to the case of pure Einstein-Maxwell theory. We shall not considered this case anymore in what follows.

 The Lagrangian \eqref{act_lam_a} can be used to study the stability of the theory. The first term in the above expression coincides with the corresponding term found in the holographic dRGT massive gravity in~\cite{Alberte:2014bua}. The equations of motion found by varying this action with respect to the fields $a$ and $\lambda$ can be easily recast in terms of the metric perturbations by expressing $\lambda$ and its derivatives from the equations \eqref{matrix} and \eqref{hfield}. The equations of motion that follow from the action \eqref{act_lam_a} are:
\begin{align}\label{eom1}
&\left(fa'\right)'-\frac{\ddot a}{f}-\frac{2\mu^2r^2}{r_h^2}a+\frac{2r^2\mu}{r_h}\lambda=0 \; ,\\\label{eom2}
&-2m_1^2\frac{f^{\alpha-2}}{M^2(r)}\ddot\lambda+m_8^2\frac{1}{M(r)}\left(\frac{\dot\lambda}{M(r)}\right)'-m_7^2\left(\frac{f^{2-\alpha}}{M^2(r)}\lambda'\right)'+\frac{r^2\det\mathcal P}{2L^2}\left(\frac{\mu}{r_h}a-\lambda\right)=0 \; .
\end{align}

In appendix~\ref{sec:stab} we analyse the stability of the vector modes in the high-frequency approximation. We find that the local causal structure for the field $\lambda$ is determined by an effective metric $\widetilde g_{ab}$ given in \eqref{eff_metric2}. In general, this emergent light cone is different from the light cone determined by the background metric $\hat g_{ab}$. In particular, if the effective light cone is wider than that of the Reissner-Nordstr\"om background, the perturbations $\lambda$ propagate superluminally with respect to $\hat g_{ab}$. A similar behaviour can be seen when discussing the perturbations of the k-essence~\cite{ArmendarizPicon:2005nz,Babichev:2007dw} and Galileons~\cite{Nicolis:2008in}. For the emergent geometry to be causally stable it has to satisfy several consistency requirements investigated in detail in appendix~\ref{sec:stab}.

\subsection{Scalar modes}\label{sec:scalars}
We further consider the stability of the homogeneous scalar sector modes~\eqref{scalar}. The relevant mass terms include
\begin{align}\label{mass_scalar}
\mathcal L_\phi (H^{AB},\phi^r)&= \frac{1}{2}\left ( m_{0}^2(\phi^r) (H^{t t})^2 - m_2^2(\phi^r) H^{i j} H^{i j} + m_{3}^2(\phi^r) H^{i i} H^{j j} \right. \notag \\
&\quad   -2m_{4}^2(\phi^r) H^{t t} H^{i i} + m_{5}^2(\phi^r) (H^{r r})^2 + m_{6}^2(\phi^r) H^{t t} H^{r r } \\
&\quad   +m_{9}^2(\phi^r) H^{r r} H^{i i}+ m_{10}^2(\phi^r)H^{rt}H^{rt} \notag\\
&\quad  +\big.m_{11}^2(\phi^r)H^{rt}H^{ii}+m_{12}^2(\phi^r)H^{tt}H^{rt}+m_{13}^2(\phi^r)H^{rr}H^{rt}\big ) \;,\notag \end{align} 
where we allow for an arbitrary $\phi^r$ dependence of the masses.

There are in total eight components of fields which transform as scalars under the transverse rotations: $h_{tt},h_{tr},h_{rr},h,a_t,a_r,\pi^t,\pi^r$. Two combinations of them can be set to zero by an appropriate choice of the coordinate transformations $\tilde t=t+\xi^t,\,\tilde r=r+\xi^r$. Moreover, due to the $U(1)$ symmetry of the Maxwell theory the components $a_t,a_r$ enter the equations of motion only in the gauge invariant combination specified below. Hence, there should be only five independent gauge invariant combinations of the scalar perturbations. Three of them can be found to be~\cite{Alberte:2014bua}:
\begin{align}\label{sc1}
&\bar h_{r}\equiv \tilde h_{rr}+\frac{1}{f}\left(r\tilde h'-\frac{rf'}{2f}\tilde h\right),\\\label{sc2}
&\bar h_t\equiv \left(\frac{1}{f}X\right)'-\frac{2}{f}\dot Y \; , \\ \label{ainv}
&\bar a\equiv a_t'-\dot a_r+\frac{\hat A_t'}{2f}X+\hat A_{t}'\left(\frac{r}{2}\tilde h\right)',
\end{align}
where we have introduced the short hand notations
\be
X=\tilde h_{tt}+\left(f-\frac{rf'}{2}\right)\tilde h \; , \qquad Y=\tilde h_{tr}+\frac{r}{2f}\dot{\tilde h} \; ,
\ee
and
\be
\tilde h_{\mu\nu}\equiv \frac{r^2}{L^2}h_{\mu\nu} \; .
\ee
These three gauge invariant fields allow us to rewrite the Einstein-Maxwell part of the action \eqref{action} up to second order in perturbations as
\begin{equation}\label{action_sc}
\L_{\text{E-M}}=\frac{L^2}{4r^4} \left( 3f^2\bar h_r^2+\frac{2\mu r^4}{r_h}f\bar h_r\bar a+2r^4\bar a^2-2rf^2\bar h_r\bar h_t\right).
\end{equation}
As we see, in the absence of the graviton mass term, there are no dynamical degrees of freedom in the scalar sector of the Einstein-Maxwell theory. 

For writing the mass term \eqref{mass_scalar} in a diffeomorphism invariant form, it is useful to introduce other gauge invariant combinations:
\begin{align}
&\bar h \equiv \tilde h+\frac{2}{r}\pi^r \; ,\\
&\bar\pi_t\equiv-\tilde h_{tt}-2f\dot\pi^t+\left(\frac{2}{r}f-f'\right)\pi^r \; ,\\
&\bar\pi_r\equiv-\tilde h_{rr}+\frac{2}{f}(\pi^r)'-\frac{1}{f}\left(\frac{2}{r}+\frac{f'}{f}\right)\pi^r \; ,\\
&\bar\pi_{tr}\equiv \tilde h_{tr}+f(\pi^t)'-\frac{1}{f}\dot\pi^r \; .
\end{align}
In terms of these fields the different components of $H^{AB}$ in the graviton mass term read:
\begin{align}\label{HAB1}
&H^{tt}=\frac{r^2}{L^2}\frac{1}{f^2}\bar\pi_t \; ,\qquad H^{rr}=\frac{r^2}{L^2}f^2\bar\pi_r \; ,\\\label{HAB2}
&H^{tr}=\frac{r^2}{L^2}\bar\pi_{tr} \; ,\qquad H^{ij}=-\delta^{ij}\frac{r^2}{L^2}\bar h\;.
\end{align}

As explained above, there are in total only five independent gauge invariant combinations of the metric, Maxwell, and St\"uckelberg field perturbations. This means that there are two relations between the seven gauge invariant fields $\bar h,\bar h_t,\bar h_r,\bar a,\bar\pi_t,\bar\pi_r,\bar\pi_{tr}$. We find these to be:
\be\label{hrht}
\bar h_r=\frac{r}{2}\left(\frac{\bar h}{f}\right)'-\bar\pi_r \; ,\qquad\bar h_t=\left(\left(1-\frac{r}{2}\frac{f'}{f}\right)\bar h\right)'-\frac{r}{f^2}\ddot{\bar h}-\left(\frac{\bar\pi_t}{f}\right)'-\frac{2}{f}\dot{\bar\pi}_{tr} \; .
\ee
These expressions allow one to eliminate the fields $\bar h_r$ and $\bar h_t$ from the action \eqref{action_sc}. Moreover, from \eqref{hrht} we see directly how the presence of the St\"uckelberg fields $\pi^t \; ,\pi^r$ introduces dynamics in the scalar sector of the theory. The previously non-derivative terms $\bar h_t$ and $\bar h_r$ turn into terms with second order time derivatives in the action. 

We thus obtain the final second order action for perturbations in two steps. We first replace the fields $\bar h_t$ and $\bar h_r$ in the Einstein-Maxwell action \eqref{action_sc} by the corresponding equations \eqref{hrht}. We then add the mass term \eqref{mass_scalar} where we replace the different components of the matrix $H^{AB}$ by the corresponding expressions \eqref{HAB1} and \eqref{HAB2}. The resulting action depends on the five gauge invariant fields $\bar a,\bar h,\bar\pi_r,\bar\pi_t,\bar\pi_{tr}$. Three of them are non-dynamical and can be integrated out or act as Lagrange multipliers and impose constraints among the remaining fields. The non-dynamical fields are the $\bar a,\bar\pi_t,\bar\pi_{tr}$ while the two dynamical degrees of freedom are carried by $\bar h$ and $\bar\pi_r$. Hence, there are at most two propagating degrees of freedom in the scalar sector of the Einstein-Maxwell massive gravity. It is a standard result in general theories of massive gravity. Around Minkowski background it is known that one of the two scalar degrees of freedom is unhealthy and propagates a ghost. This happens in both Lorentz invariant~\cite{Fierz:1939ix,Boulware:1973my} and Lorentz violating~\cite{Rubakov:2004eb,Dubovsky:2004sg} massive gravity theories. In Lorentz violating massive gravity theories the number of degrees of freedom in the scalar sector depends on the choice of the graviton masses: by setting some of the $m_i$'s to zero, one can propagate either two, one or none of the helicity-0 fields. Below we identify the mass terms that are responsible for the degree of freedom count and present the main choices that lead to distinct number of propagating fields.

We shall not present the full Lagrangian here. The expression is quite long and becomes even longer after we start integrating out the non-dynamical fields. We list the different steps we take in manipulating the action and discuss the different results depending on the mass parameters $m_i$. The actual calculations are straightforward. The first step is to integrate out the Maxwell field $\bar a$. It enters the action linearly and quadratically without derivatives, which allows use to use its equation of motion to eliminate it from the action. The next step is to integrate out the field $\bar\pi_{tr}$. It enters the action as:
\be\notag
\mathcal L_{\bar\pi_{tr}}=\frac{1}{2}m_{10}^2(r)\bar\pi_{tr}^2-\bar\pi_{tr}\,\mathcal F_1\left(\dot{\bar h}',\dot{\bar h},\dot{\bar\pi}_r,\bar h,\bar\pi_t,\bar\pi_r\right) \; .
\ee
The result depends crucially on whether the mass parameter $m_{10}$ in front of the term quadratic in $\bar\pi_{tr}$ is vanishing or not.

\bigskip
In the case of a non-vanishing $m_{10}$ the equation of motion of $\bar\pi_{tr}$ allows to express it in terms of the other remaining fields $\bar h,\bar\pi_t,\bar\pi_r$ and eliminates it from the action. We then proceed by integrating out the field $\pi_t$ which appears in the action as
\begin{equation*}
\mathcal L_{\bar\pi_t}=\frac{1}{8m_{10}^2(r)f^{4}}\left(4m_0^2(r)m_{10}^2(r)-m_{12}^4(r)\right)\bar\pi_t^2-\bar\pi_t\mathcal F_2\left(\dot{\bar h}',\dot{\bar h},\dot{\bar\pi}_r,\bar h'',\bar h',\bar\pi_r',\bar h,\bar\pi_r\right) \; .
\end{equation*}
Further results depend on whether the combination $4m_0^2m_{10}^2-m_{12}^4$ is vanishing or not.

\begin{itemize}
\item $4m_0^2m_{10}^2-m_{12}^4\neq 0$

In this case we use the equation of motion for the field $\bar\pi_t$ to eliminate it from the action. The remaining action contains the two fields $\bar h$ and $\bar\pi_r$. They both appear with quadratic time derivatives and are kinetically mixed. We thus see two propagating degrees of freedom. In analogy with the usual behaviour of the helicity-0 fields in massive gravity we expect that one of them is a ghost. We therefore conclude, that the massive gravity theory \eqref{action} with the mass term \eqref{mass_scalar} is unstable in this case. 

\item $4m_0^2m_{10}^2-m_{12}^4= 0$

In this case the quadratic term $\bar\pi_t^2$ vanishes from the action and the field $\bar\pi_t$ becomes a Lagrange multiplier imposing the constraint $\mathcal F_2=0$ on the fields $\bar h$ and $\bar\pi_r$. This particular constraint allows to eliminate all the spatial derivatives of $\bar\pi_r$ from the action. It means that in a high frequency limit one of the degrees of freedom has the dispersion relation $\omega^2=0$ and does not propagate. The other degree of freedom is a linear combination of the fields $\bar h',\,\bar h,\,\bar\pi_r$ and is dynamical. Hence, we conclude that there is one propagating helicity-0 degree of freedom in the high energy limit and this phase can be stable.

\end{itemize}

\bigskip
When the mass parameter $m_{10}$ vanishes, the field $\bar\pi_{tr}$ turns into a Lagrange multiplier and its equation of motion imposes a constraint $\mathcal F_1$ on the three fields $\bar h,\bar\pi_t,\bar\pi_r$. The constraint has the form
\begin{equation}
\mathcal F_{1}=-\frac{1}{2f^2}m_{12}^2(r)\bar\pi_t+\mathcal O\left(\dot{\bar h}',\dot{\bar h},\dot{\bar\pi}_r,\bar h,\bar\pi_r\right)=0 \;.
\end{equation}
Further results depend on the choice of $m_{12}$.

\begin{itemize}
\item $m_{12}\neq 0$

For non-zero mass parameter $m_{12}$ we can use the constraint $\mathcal F_1=0$ to eliminate $\bar\pi_t$ from the action. The resulting action contains second order time derivatives of the both remaining fields $\bar h$ and $\bar\pi_r$. This shows that there are two propagating degrees of freedom. As before, we conclude that this case is unstable since one of the degrees of freedom is a ghost.

\item $m_{12}=0$ and $m_0\neq 0$

In this case we cannot use the constraint $\mathcal F_1$ to eliminate the field $\bar\pi_t$. Instead the constraint provides a relation between the fields $\bar h$ and $\bar\pi_r$. We use it in the action in order to remove the term $\bar\pi_{tr}\mathcal F_1$. The field $\bar\pi_t$ enters the remaining action only linearly and quadratically without derivatives and can thus be integrated out by its equation of motion. In the final action there are no second order time derivatives. We conclude that there are no propagating helicity-0 degrees of freedom in this case.

\item $m_{12}=0$ and $m_0=0$

As before we use the constraint $\mathcal F_1$ to remove the term $\bar\pi_{tr}$ from the action. We further use it to express the spatial derivative $\bar\pi_r'$ and eliminate it from the action. As a result the only remaining terms that involve the field $\bar\pi_r$ are: $\dot{\bar\pi}_r\dot{\bar h},\,\bar\pi_r^2,\,\bar\pi_r\bar h$. The highest derivative term of the $\bar h$ field that arises in the quadratic action is $(\dot{\bar h})^2,\,(\bar h')^2$. In the high frequency limit this leads to the fact that both fields have a dispersion relation $\omega^2=0$. We thus conclude that there are no propagating degrees of freedom in this case. 

\end{itemize}

\subsubsection*{Conclusion about the stability}
To summarise we conclude that the theory propagates two degrees of freedom and is thus necessarily unstable if 
\begin{equation}
4m_0^2m_{10}^2-m_{12}^4\neq 0\; .
\end{equation}
The high energy limit allows us to further conclude that there is at most one dynamical scalar degree of freedom in the parameter region:
\begin{equation}\label{high_en1}
4m_0^2m_{10}^2-m_{12}^4= 0 \;, \qquad\text{and}\qquad m_{10} \neq 0 \;.
\end{equation}
Finally, we find that there are no dynamical helicity-0 degrees of freedom if
\begin{equation}
m_{10}=0 \;, \qquad \text{and} \qquad m_{12} = 0 \;. \label{high_en2}
\end{equation}
We note that from the above cases, \eqref{high_en1} and \eqref{high_en2} were obtained in the high frequency limit. 

Our results presented here coincide with the earlier results for the Lorentz violating massive gravity around the Minkowski background in~\cite{Rubakov:2004eb,Dubovsky:2004sg}. In particular, in the high energy limit it was found that there is one propagating scalar degree of freedom when $m_0^2=0,\,m_{10}^2\neq 0$ and no propagating scalars when $m_{10}^2=0$, wherease the case with $m_0^2\neq 0$ and $m_{10}^2\neq 0$ is unstable~\cite{Dubovsky:2004sg}\footnote{We have identified the mass term $m_1$ of~\cite{Dubovsky:2004sg} with the mass term $m_{10}$ in our case.} . These findings coincide with our conclusions above.

\section{Stability analysis in eikonal approximation}\label{sec:stab}
Here we investigate the short wavelength dynamics of the helicity-1 fields $a$ and $\lambda$ that describe the dynamics of the vector perturbations of the Maxwell field and the metric and were introduced in appendix~\ref{sec:vectors}. We analyze the action \eqref{act_lam_a} and the corresponding equations of motion \eqref{eom1} and \eqref{eom2} by using the eikonal approximation~\cite{landau2}. For this we expand the fields as $a(t,r)\propto\exp(i\omega \hat S(t,r))$ and $\lambda(t,r)\propto\exp(i\omega S(t,r))$ and take the high frequency limit $\omega\to\infty$. In this limit only the terms quadratic in $\omega$ survive and the equations read
\be\label{eff_metric}
\hat g^{ab}\pt_a\hat S\,\pt_b\hat S=0 \; ,\qquad\text{and}\qquad \widetilde g^{ab}\pt_aS\,\pt_bS=0 \; ,
\ee
where the indices $a,b=\{t,r\}$ and $\hat g^{ab}$ is the background metric \eqref{solmetric} defining the metric on which the perturbations of the Maxwell field propagate. In turn, the metric $\widetilde g^{ab}$ is a new effective metric defining the geometry on which the perturbations $\lambda$ propagate:
\be\label{eff_metric2}
\widetilde g^{ab}=
\begin{pmatrix}
-2m_1^2f^{\alpha-2}&\frac{1}{2}m_8^2\\
\frac{1}{2}m_8^2&-m_7^2f^{2-\alpha}
\end{pmatrix}\;.
\ee
The physical meaning of the equations \eqref{eff_metric} is that at the leading order in the eikonal limit the fields $a$ and $\lambda$ can be approximated as waves with a slowly changing phase $\hat S$ and $ S$ respectively. The surface of constant phase is the corresponding wavefront and its normal is given by the gradient $\pt_a S$. The effective acoustic cone of the wavefront propagation is determined by the vectors tangential to the wavefront $n^a\equiv\widetilde g^{ab}\pt_b S$. That the vector $n^a$ is indeed tangential to the characteristic surface follows from equation \eqref{eff_metric}. By further rewriting of the equation \eqref{eff_metric} we find another equation 
\be
\widetilde g_{ab}n^an^b=0\;,
\ee
where $\widetilde g_{ab}$ is the inverse of $\widetilde g^{ab}$. Hence, the wavefronts of the field $\lambda$ moves as if it were light in the spacetime $\widetilde g_{ab}$. 

The effective metric $\widetilde g_{ab}$ arises due to the scalar field condensate that fills the Reissner-Nordstr\"om spacetime and defines the local causal structure for the field $\lambda$. In general, this emergent light cone is different from the light cone determined by the background metric $\hat g_{ab}$. In particular, if the effective light cone is wider than that of the Reissner-Nordstr\"om background, the perturbations $\lambda$ propagate superluminally with respect to $\hat g_{ab}$. In such situation, there are several conditions the emergent geometry has to satisfy in order for it to be causally stable. 

\subsubsection*{Hyperbolicity}
Obviously, the equations of motion \eqref{eff_metric} have to by hyperbollic, i.e. the metric $\tilde g^{ab}$ has to have a Lorentzian signature:
\be\label{condition1}
\det \widetilde g^{-1}=2m_1^2m_7^2-\frac{1}{4}m_8^4=-\frac{1}{4}\det\mathcal P<0 \; .
\ee
This means that $\det\mathcal P=m^4_8-8m_1^2m_7^2>0$ and imposes the first condition on the graviton masses $m_i^2$. 

\subsubsection*{Closed time-like curves}
Due to the fact that, in general, superluminal propagation might be possible in the metric $\widetilde g^{ab}$ one has to investigate whether closed time-like curves can possibly form. To prove the causal stability of the emergent metric $\widetilde g_{ab}$ it is sufficient to find a differentiable function $\tau$ so that $\partial_\mu \tau$ is a future directed time-like vector field (see~\cite{Babichev:2007dw,wald} and references therein). The function $\tau$ then serves the role of a global time function of the spacetime under consideration. We choose the function $\tau$ to be the time $t$ of the original Reissner-Nordstr\"om spacetime. Then the following condition has to be satisfied
\be\label{condition2}
\widetilde g^{ab}\pt_at\,\pt_bt=\widetilde g^{tt}=-2m_1^2f^{\alpha-2}<0\;.
\ee
Depending on the choice of $\alpha$ this expression can become infinite on the horizon when $f(r_h)=0$ stressing the fact that the coordinate $t$ is not a good choice for the time direction near the horizon. This is the case also for the Reissner-Nordstr\"om metric and thus we will disregard this issue here. The above condition together with \eqref{condition1} implies the following constraints on the masses:
\be\label{condition3}
m_1^2>0\;,\qquad \text{and}\qquad m_7^2<m_8^4/(8m_1^2)\;.
\ee
In the special case when $m_8^2=0$ this leads to $m_7^2<0$.

\subsubsection*{Light cones}

In order, to compare the light cone defined by $\widetilde g_{ab}$ with the one defined by $\hat g_{ab}$ we consider a vector $v^a$ that is null with respect to the metric $\widetilde g$, i.e. $\widetilde g_{ab}v^av^b=0$. We then investigate whether this vector is space-like or time-like with respect to the background metric $\hat g_{ab}$. By knowing that the vector $v^a$ is a null vector with respect to $\widetilde g$ we find
\be
v^r_\pm=\frac{1}{2}\frac{f^{2-\alpha}v^t}{2m_1^2}\left(-m_8^2\pm\sqrt{\det\mathcal P}\right) \; ,
\ee
where the plus and minus indicate the two boundaries of the light cone. It is interesting to notice, that only in the case $m_8^2=0$ the light cone of $\widetilde g$ is symmetric with respect to the time axis. One can further check that 
\be
\hat g_{ab}v^a_\pm v^b_\pm=\frac{L^2}{r^2}\frac{f(v^t)^2}{(4m_1^2)^2}\left[f^{2(1-\alpha)}\left(-m_8^2\pm\sqrt{\det\mathcal P}\right)^2-(4m_1^2)^2\right] \; .
\ee
The two light cones coincide only if both $\hat g_{ab}v^a_+v^b_+=\hat g_{ab}v^a_-v^b_-=0$. The sign of $\hat g_{ab}v^a_\pm v^b_\pm$ determines whether the light cone of $\widetilde g$ is wider or narrower than the light cone of $\hat g$. We note that the mutual alignment of the two light cones for some fixed values of the masses $m_i^2$ can flip while the radial coordinate changes from $0$ to $r_h$ due to the factor of $f^{2(1-\alpha)}$ inside the square brackets. In order for this not to happen we demand that 
\be\label{cond_alpha2}
\alpha=1\;.
\ee
Under the above assumption the effective metric simplifies to
 \be\label{eff_metric3}
 \widetilde g^{ab}=
\begin{pmatrix}
-2m_1^2f^{-1}&\frac{1}{2}m_8^2\\
\frac{1}{2}m_8^2&-m_7^2f^{}
\end{pmatrix}\;.
\ee

A general condition for the two metrics $\hat g_{ab}$ and $\widetilde g_{ab}$ to have the same causal structure is that they are related to each other by a conformal transformation $\Omega(r)$ so that $\widetilde g_{ab}=\Omega\hat g_{ab}$. Since the light cone of $\hat g_{ab}$ is symmetric with respect to the time axis, i.e. $\hat g_{ab}$ is a diagonal metric, then this is possible only when $m_8^2=0$. The condition $\hat g_{ab}v^a_\pm v^b_\pm=0$ then simplifies to
\be\label{condition4}
m_7^2+2m_1^2=0\;,\qquad\text{when}\qquad m_8^2=0\;.
\ee
We also note that on the horizon $f(r_h)=0$ and hence both light cones coincide there independently on the choice of the graviton masses $m_i^2$.  

Other particular choices of the graviton masses $m_i^2$ correspond to the cases when one of the sides of the light cone of $\widetilde g$ coincides with a side of the light cone of $\hat g$. This happens when one of the conditions $\hat g_{ab}v^a_+v^b_+=0$ and $\hat g_{ab}v^a_-v^b_-=0$ is satisfied. For $v^a_+$ this occurs when
\be
\begin{cases}\label{special_masses}
m_8^2=-m_7^2-2m_1^2\;,\qquad&\text{when}\qquad m_7^2\leq2m_1^2\;,\\
m_8^2=m_7^2+2m_1^2\;,\quad&\text{when}\qquad m_7^2\geq2m_1^2\;.\\
\end{cases}
\ee
Similarly, for $v^a_-$ we obtain
\be
\begin{cases}\label{special_masses0}
m_8^2=-m_7^2-2m_1^2\;,\qquad&\text{when}\qquad m_7^2\geq2m_1^2\;,\\
m_8^2=m_7^2+2m_1^2\;,\quad&\text{when}\qquad m_7^2\leq2m_1^2\;.\\
\end{cases}
\ee

To analyse the mutual alignment of the two light cones for a particular choice of the masses $m_i^2$ we have to check the sign of the norms $g_{ab}v^a_\pm v^b_\pm$. If both $g_{ab}v^a_\pm v^b_\pm>0$, then the null vectors  $v^a_\pm$ with respect to the metric $\widetilde g_{ab}$ are spacelike with respect to the Reissner-Nordstr\"om metric $\hat g_{ab}$ meaning that the light cone determined by the new effective metric $\tilde g$ is wider than the light cone of $\hat g$. Hence, the perturbations $\lambda$ can propagate superluminally. 

\subsubsection*{Summary}
From the comparison of the two light cones defined by the background metric and the effective metric, we first conclude that for the mutual alignment of the two cones not to flip while the radial coordinate changes from $0$ to $r_h$ we need to demand that $\alpha=1$. We further find that a necessary condition for the two light cones to coincide is given in \eqref{condition4} whereas the requirements for only one of the sides of the light cones to coincide read:
\be
\begin{cases}\label{special_masses_0}
&m_8^2=-m_7^2-2m_1^2\;,\qquad\text{or}\\
&m_8^2=m_7^2+2m_1^2\;.\\
\end{cases}
\ee

In general, whenever $m_8^2\neq 0$ superluminal propagation of the perturbations $\lambda$ becomes possible. This result has been obtained in the eikonal approxmation that corresponds to taking the high frequency limit. In this limit, the phase velocity determined by the effective metric $\widetilde g$ approaches the front velocity of the propagation of $\lambda$ and describes the speed of the signal propagation. Such a situation for perturbations propagating on a Minkowski background would be pathological and should be avoided. However, as has been argued previously in~\cite{Babichev:2007dw}, having a superluminal propagation of perturbations on an emergent effective geometry due to a condensate of scalar fields does not lead to any causal paradoxes as long as certain conditions are met. These include the requirement of the causal stability of the emergent spacetime and of a well-posed Cauchy problem on an initial hypersurface that is spacelike with respect to both the effective metric $\widetilde g$ and the gravitational metric $\hat g$~\cite{Babichev:2007dw}. From our analysis above we find that, the spacetime defined by $\widetilde g_{ab}$ is causally stable if $m_1^2>0$ whereas the hyperbolicity condition is satisfied when $m_7^2<m_8^4/(8m_1^2)$. The issue of correctly posed Cauchy problem is more involved and has not be discussed here. 

We also note that, strictly speaking, the eikonal approximation assumes that the underlying theory is UV complete. The theory at hand is, however, an effective theory and is known to have a very low UV cutoff scale~\cite{ArkaniHamed:2002sp}. It might therefore be that the superluminal propagation found in the eikonal limit is an artifact of the high frequency limit and would not be present in the hypothethic UV complete massive gravity theory (for a recent proposal of such a theory, see~\cite{Blas:2014ira}). Whether or not such a completion is possible in the presence of superluminalities in the low energy effective field theory is another subtle question that has been studied in a number of earlier works~\cite{Adams:2006sv,Keltner:2015xda} and is beyond the scope of this work.
 
We therefore conclude that restricting the allowed range of the graviton masses $m_i^2$ basing on the fact that the theory permits superluminal propagation of field perturbations would be premature. Instead we shall accept the point of view that a superluminality in an effective field theory is allowed and investigate whether this causes any inconsistencies in the dual field theory.


\begin{thebibliography}{2}   



  
    
  \bibitem{Vegh:2013sk}
  D.~Vegh,
  ``Holography without translational symmetry,''
  arXiv:1301.0537 [hep-th].
  
\bibitem{Blake:2013bqa}
  M.~Blake and D.~Tong,
  ``Universal Resistivity from Holographic Massive Gravity,''
  Phys.\ Rev.\ D {\bf 88} (2013) 10,  106004
  [arXiv:1308.4970 [hep-th]].
  
  \bibitem{Blake:2013owa}
  M.~Blake, D.~Tong and D.~Vegh,
  ``Holographic Lattices Give the Graviton an Effective Mass,''
  Phys.\ Rev.\ Lett.\  {\bf 112} (2014) 7,  071602
  [arXiv:1310.3832 [hep-th]].
  
    \bibitem{Davison:2013jba}
  R.~A.~Davison,
  ``Momentum relaxation in holographic massive gravity,''
  Phys.\ Rev.\ D {\bf 88} (2013) 086003
  [arXiv:1306.5792 [hep-th]].

  \bibitem{Rubakov:2004eb}
  V.~A.~Rubakov,
  ``Lorentz-violating graviton masses: Getting around ghosts, low strong coupling scale and VDVZ discontinuity,''
  hep-th/0407104.

\bibitem{Dubovsky:2004sg}
  S.~L.~Dubovsky,
  ``Phases of massive gravity,''
  JHEP {\bf 0410} (2004) 076
  [hep-th/0409124].
  
  \bibitem{Blas:2014ira} 
  D.~Blas and S.~Sibiryakov,
  ``Completing Lorentz violating massive gravity at high energies,''
  Zh.\ Eksp.\ Teor.\ Fiz.\  {\bf 147}, 578 (2015)
  [J.\ Exp.\ Theor.\ Phys.\  {\bf 120}, no. 3, 509 (2015)]
  [arXiv:1410.2408 [hep-th]].
  
   \bibitem{deRham:2014zqa}
  C.~de Rham,
  ``Massive Gravity,''
  Living Rev.\ Rel.\  {\bf 17} (2014) 7
  [arXiv:1401.4173 [hep-th]].
  
  \bibitem{Maldacena:1997re}
  J.~M.~Maldacena,
  ``The Large N limit of superconformal field theories and supergravity,''
  Int.\ J.\ Theor.\ Phys.\  {\bf 38} (1999) 1113
   [Adv.\ Theor.\ Math.\ Phys.\  {\bf 2} (1998) 231]
  [hep-th/9711200].

    \bibitem{deRham:2010kj}
  C.~de Rham, G.~Gabadadze and A.~J.~Tolley,
  ``Resummation of Massive Gravity,''
  Phys.\ Rev.\ Lett.\  {\bf 106} (2011) 231101
  [arXiv:1011.1232 [hep-th]].
  
\bibitem{Baggioli:2014roa} 
  M.~Baggioli and O.~Pujolas,
  ``Electron-Phonon Interactions, Metal-Insulator Transitions, and Holographic Massive Gravity,''
  Phys.\ Rev.\ Lett.\  {\bf 114}, no. 25, 251602 (2015)
  [arXiv:1411.1003 [hep-th]].
  
  \bibitem{Alberte:2014bua}
  L.~Alberte and A.~Khmelnitsky,
  ``Stability of Massive Gravity Solutions for Holographic Conductivity,''
  Phys.\ Rev.\ D {\bf 91} (2015) 4,  046006
  [arXiv:1411.3027 [hep-th]].
  
  
    \bibitem{Andrade:2013gsa}
  T.~Andrade and B.~Withers,
  ``A simple holographic model of momentum relaxation,''
  JHEP {\bf 1405} (2014) 101
  [arXiv:1311.5157 [hep-th]].
  
  \bibitem{Taylor:2014tka}
  M.~Taylor and W.~Woodhead,
  ``Inhomogeneity simplified,''
  Eur.\ Phys.\ J.\ C {\bf 74} (2014) 12,  3176
  [arXiv:1406.4870 [hep-th]].
  
\bibitem{Leutwyler:1993gf} 
  H.~Leutwyler,
  ``Nonrelativistic effective Lagrangians,''
  Phys.\ Rev.\ D {\bf 49}, 3033 (1994);
  [hep-ph/9311264].

  \bibitem{Leutwyler:1996er} 
  H.~Leutwyler,
  ``Phonons as goldstone bosons,''
  Helv.\ Phys.\ Acta {\bf 70}, 275 (1997)
  [hep-ph/9609466].
  
  

\bibitem{Nicolis:2015sra} 
  A.~Nicolis, R.~Penco, F.~Piazza and R.~Rattazzi,
  ``Zoology of condensed matter: Framids, ordinary stuff, extra-ordinary stuff,''
  JHEP {\bf 1506}, 155 (2015)
  [arXiv:1501.03845 [hep-th]].

\bibitem{Nicolis:2013lma} 
  A.~Nicolis, R.~Penco and R.~A.~Rosen,
  ``Relativistic Fluids, Superfluids, Solids and Supersolids from a Coset Construction,''
  Phys.\ Rev.\ D {\bf 89}, no. 4, 045002 (2014)
  [arXiv:1307.0517 [hep-th]].

\bibitem{Dubovsky:2011sj}
  S.~Dubovsky, L.~Hui, A.~Nicolis and D.~T.~Son,
  ``Effective field theory for hydrodynamics: thermodynamics, and the derivative expansion,''
  Phys.\ Rev.\ D {\bf 85} (2012) 085029
  [arXiv:1107.0731 [hep-th]].

\bibitem{Son:2005ak}
  D.~T.~Son,
  ``Effective Lagrangian and topological interactions in supersolids,''
  Phys.\ Rev.\ Lett.\  {\bf 94} (2005) 175301
  [cond-mat/0501658].
  
  \bibitem{Endlich:2012pz}
  S.~Endlich, A.~Nicolis and J.~Wang,
  ``Solid Inflation,''
  JCAP {\bf 1310} (2013) 011
  [arXiv:1210.0569 [hep-th]].
  

\bibitem{Hartnoll:2009sz}
  S.~A.~Hartnoll,
  ``Lectures on holographic methods for condensed matter physics,''
  Class.\ Quant.\ Grav.\  {\bf 26} (2009) 224002
  [arXiv:0903.3246 [hep-th]].

  
  
\bibitem{Hartnoll:2008hs} 
  S.~A.~Hartnoll and C.~P.~Herzog,
  ``Impure AdS/CFT correspondence,''
  Phys.\ Rev.\ D {\bf 77}, 106009 (2008)
  [arXiv:0801.1693 [hep-th]].

  
\bibitem{Iqbal:2008by}
  N.~Iqbal and H.~Liu,
  ``Universality of the hydrodynamic limit in AdS/CFT and the membrane paradigm,''
  Phys.\ Rev.\ D {\bf 79} (2009) 025023
  [arXiv:0809.3808 [hep-th]].

\bibitem{Hartnoll:2009ns} 
  S.~A.~Hartnoll, J.~Polchinski, E.~Silverstein and D.~Tong,
  ``Towards strange metallic holography,''
  JHEP {\bf 1004}, 120 (2010)
  [arXiv:0912.1061 [hep-th]].


\bibitem{Davison:2013txa}
  R.~A.~Davison, K.~Schalm and J.~Zaanen,
  ``Holographic duality and the resistivity of strange metals,''
  Phys.\ Rev.\ B {\bf 89} (2014) 24,  245116
  [arXiv:1311.2451 [hep-th]].




\bibitem{Hartnoll:2014lpa} 
  S.~A.~Hartnoll,
  ``Theory of universal incoherent metallic transport,''
  Nature Phys.\  {\bf 11}, 54 (2015)
  [arXiv:1405.3651 [cond-mat.str-el]].


\bibitem{Donos:2014uba}
  A.~Donos and J.~P.~Gauntlett,
  ``Novel metals and insulators from holography,''
  JHEP {\bf 1406} (2014) 007
  [arXiv:1401.5077 [hep-th]].
  
\bibitem{Donos:2014cya}
  A.~Donos and J.~P.~Gauntlett,
  ``Thermoelectric DC conductivities from black hole horizons,''
  JHEP {\bf 1411} (2014) 081
  [arXiv:1406.4742 [hep-th]].
  
\bibitem{Amoretti:2014ola}
  A.~Amoretti, A.~Braggio, N.~Magnoli and D.~Musso,
  ``Bounds on charge and heat diffusivities in momentum dissipating holography,''
  JHEP {\bf 1507} (2015) 102
  [arXiv:1411.6631 [hep-th]].

  
\bibitem{Grozdanov:2015qia} 
  S.~Grozdanov, A.~Lucas, S.~Sachdev and K.~Schalm,
  ``Absence of disorder-driven metal-insulator transitions in simple holographic models,''
  arXiv:1507.00003 [hep-th].
  
\bibitem{Davison:2014lua} 
  R.~A.~Davison and B.~Gouteraux,
  ``Momentum dissipation and effective theories of coherent and incoherent transport,''
  JHEP {\bf 1501}, 039 (2015)
  [arXiv:1411.1062 [hep-th]].

  

\bibitem{Hartnoll:2014cua} 
  S.~A.~Hartnoll and J.~E.~Santos,
  ``Disordered horizons: Holography of randomly disordered fixed points,''
  Phys.\ Rev.\ Lett.\  {\bf 112}, 231601 (2014)
  [arXiv:1402.0872 [hep-th]].



\bibitem{Gouteraux:2014hca} 
  B.~Goutraux,
  ``Charge transport in holography with momentum dissipation,''
  JHEP {\bf 1404}, 181 (2014)
  [arXiv:1401.5436 [hep-th]].

\bibitem{Hartnoll:2015rza} 
  S.~A.~Hartnoll, D.~M.~Ramirez and J.~E.~Santos,
  ``Thermal conductivity at a disordered quantum critical point,''
  arXiv:1508.04435 [hep-th].
  
\bibitem{Lucas:2015lna} 
  A.~Lucas,
  ``Hydrodynamic transport in strongly coupled disordered quantum field theories,''
  arXiv:1506.02662 [hep-th].

\bibitem{Lucas:2014sba} 
  A.~Lucas and S.~Sachdev,
  ``Conductivity of weakly disordered strange metals: from conformal to hyperscaling-violating regimes,''
  Nucl.\ Phys.\ B {\bf 892}, 239 (2015)
  [arXiv:1411.3331 [hep-th]].

\bibitem{Lucas:2014zea} 
  A.~Lucas, S.~Sachdev and K.~Schalm,
  ``Scale-invariant hyperscaling-violating holographic theories and the resistivity of strange metals with random-field disorder,''
  Phys.\ Rev.\ D {\bf 89}, no. 6, 066018 (2014)
  [arXiv:1401.7993 [hep-th]].
  
\bibitem{Arean:2015sqa} 
  D.~Arean, L.~A.~Pando Zayas, I.~S.~Landea and A.~Scardicchio,
  ``The Holographic Disorder-Driven Supeconductor-Metal Transition,''
  arXiv:1507.02280 [hep-th].

\bibitem{Arean:2013mta} 
  D.~Arean, A.~Farahi, L.~A.~Pando Zayas, I.~S.~Landea and A.~Scardicchio,
  ``Holographic superconductor with disorder,''
  Phys.\ Rev.\ D {\bf 89}, no. 10, 106003 (2014)
  [arXiv:1308.1920 [hep-th]].
  
\bibitem{Amoretti:2015gna} 
  A.~Amoretti and D.~Musso,
  ``Magneto-transport from momentum dissipating holography,''
  JHEP {\bf 1509}, 094 (2015)
  [arXiv:1502.02631 [hep-th]].
  
\bibitem{Baggioli:2015dwa} 
  M.~Baggioli and M.~Goykhman,
  ``Under The Dome: Doped holographic superconductors with broken translational symmetry,''
  arXiv:1510.06363 [hep-th].

\bibitem{Baggioli:2015zoa} 
  M.~Baggioli and M.~Goykhman,
  ``Phases of holographic superconductors with broken translational symmetry,''
  JHEP {\bf 2015}, 35
  [arXiv:1504.05561 [hep-th]].
  
\bibitem{Baggioli:2015gsa} 
  M.~Baggioli and D.~K.~Brattan,
  ``Drag Phenomena from Holographic Massive Gravity,''
  arXiv:1504.07635 [hep-th].
  
 \bibitem{deHaro:2000xn}
  S.~de Haro, S.~N.~Solodukhin and K.~Skenderis,
  ``Holographic reconstruction of space-time and renormalization in the AdS / CFT correspondence,''
  Commun.\ Math.\ Phys.\  {\bf 217} (2001) 595
  [hep-th/0002230].  
  
  \bibitem{Babichev:2007dw}
  E.~Babichev, V.~Mukhanov and A.~Vikman,
  ``k-Essence, superluminal propagation, causality and emergent geometry,''
  JHEP {\bf 0802} (2008) 101
  [arXiv:0708.0561 [hep-th]].    
    
\bibitem{Megias:2014iwa} 
  E.~Megias and O.~Pujolas,
  ``Naturally light dilatons from nearly marginal deformations,''
  JHEP {\bf 1408}, 081 (2014)
  [arXiv:1401.4998 [hep-th]].
  
\bibitem{Argurio:2014rja} 
  R.~Argurio, D.~Musso and D.~Redigolo,
  ``Anatomy of new SUSY breaking holographic RG flows,''
  JHEP {\bf 1503}, 086 (2015)
  [arXiv:1411.2658 [hep-th]].
  
  \bibitem{polaron}
J.~T.~ Devreese, ``Fr\"ohlich polaron concept. Lecture course including detailed theoretical derivations," (2010) [arXiv:1012.4576].
  
  \bibitem{landau7} L.~D.~Landau, E.~M.~Lifshitz, ``Course of Theoretical Physics,'' Vol. 7, ``Theory of Elasticity,'' 1-12, Pergamon Press (1970).

\bibitem{Lubensky} P.~M.~Chaikin, T.~C.~Lubensky, ``Principles of Condensed Matter Physics,'' Sections 6.4 and 8.4, Cambridge University Press (1995).
  
   \bibitem{future}
  L.~Alberte, M.~Baggioli, A.~Khmelnitsky, O.~Pujolas, in preparation.
  
  \bibitem{Policastro:2001yc}
  G.~Policastro, D.~T.~Son and A.~O.~Starinets,
  ``The Shear viscosity of strongly coupled N=4 supersymmetric Yang-Mills plasma,''
  Phys.\ Rev.\ Lett.\  {\bf 87} (2001) 081601
  [hep-th/0104066].
  
\bibitem{Endlich:2010hf}
  S.~Endlich, A.~Nicolis, R.~Rattazzi and J.~Wang,
  ``The Quantum mechanics of perfect fluids,''
  JHEP {\bf 1104} (2011) 102
  [arXiv:1011.6396 [hep-th]].
  
   \bibitem{Boulware:1973my}
  D.~G.~Boulware and S.~Deser,
  ``Can gravitation have a finite range?,''
  Phys.\ Rev.\ D {\bf 6} (1972) 3368.
  
 \bibitem{Alberte:2013sma}
  L.~Alberte and A.~Khmelnitsky,
  ``Reduced massive gravity with two Stückelberg fields,''
  Phys.\ Rev.\ D {\bf 88} (2013) 6,  064053
  [arXiv:1303.4958 [hep-th]].
  
    
  \bibitem{ArmendarizPicon:2005nz}
  C.~Armendariz-Picon and E.~A.~Lim,
  ``Haloes of k-essence,''
  JCAP {\bf 0508} (2005) 007
  [astro-ph/0505207].

 
  
  \bibitem{Nicolis:2008in}
  A.~Nicolis, R.~Rattazzi and E.~Trincherini,
  ``The Galileon as a local modification of gravity,''
  Phys.\ Rev.\ D {\bf 79} (2009) 064036
  [arXiv:0811.2197 [hep-th]].

\bibitem{Fierz:1939ix}
  M.~Fierz and W.~Pauli,
  ``On relativistic wave equations for particles of arbitrary spin in an electromagnetic field,''
  Proc.\ Roy.\ Soc.\ Lond.\ A {\bf 173} (1939) 211.

    \bibitem{landau2}
  L.~D.~Landau, E.~M.~Lifshitz, ``Course of Theoretical Physics,'' Vol. 2, ``The Classical Theory of Fields,'' 141-143, Butterworth, Heinemann (1994).
  
    \bibitem{wald}
  R.~Wald, ``General Relativity,'' The University of Chichago Press, Sec. 8.2 (1984).
  
  

  
  



  

  
\bibitem{ArkaniHamed:2002sp}
  N.~Arkani-Hamed, H.~Georgi and M.~D.~Schwartz,
  ``Effective field theory for massive gravitons and gravity in theory space,''
  Annals Phys.\  {\bf 305} (2003) 96
  [hep-th/0210184].
  
\bibitem{Adams:2006sv}
  A.~Adams, N.~Arkani-Hamed, S.~Dubovsky, A.~Nicolis and R.~Rattazzi,
  ``Causality, analyticity and an IR obstruction to UV completion,''
  JHEP {\bf 0610} (2006) 014
  [hep-th/0602178].
  
\bibitem{Keltner:2015xda}
  L.~Keltner and A.~J.~Tolley,
  ``UV properties of Galileons: Spectral Densities,''
  arXiv:1502.05706 [hep-th].

  
  
  
  










  
\end{thebibliography}
\end{document}